\documentclass[letterpaper,twocolumn,10pt]{article}
\usepackage{usenix-2020-09}

\PassOptionsToPackage{dvipsnames}{xcolor}
\usepackage{tikz}
\usepackage{amsmath}
\usepackage{xcolor}
\usepackage{xspace}
\usepackage{amssymb}   %
\usepackage{graphicx}
\usepackage{subcaption} %
\usepackage{booktabs}   %
\usepackage{array}      %
\usepackage{enumitem}   %
\usepackage{xstring}    %
\usepackage{algorithm}       %
\usepackage{algpseudocode}
\usepackage{cleveref}       %

\ifdefined\macrosloaded\else
\newcommand{\macrosloaded}{}

\newcommand{\ie}{i.e.,}

\newcommand{\sys}{E4\xspace}

\newcommand{\moce}{MoCE\xspace}

\newcommand{\openrca}{OpenRCA\xspace}
\newcommand{\market}{Market-CB1\xspace}
\newcommand{\bank}{Bank\xspace}
\newcommand{\telecom}{Telecom\xspace}
\newcommand{\terminus}{Terminus-2\xspace}

\newcommand{\sage}{Sage\xspace}
\newcommand{\murphy}{Murphy\xspace}

\newcommand{\tabmark}[1]{\raisebox{0pt}[1.7ex][1.2ex]{#1}}
\newcommand{\cmark}{\tabmark{$\checkmark$}}
\newcommand{\xmark}{\tabmark{X}}

\newcommand{\structuredimagination}{constrained creativity\xspace} 
\newcommand{\eespectrum}{expressivity--explainability spectrum\xspace}
\newcommand{\Structuredimagination}{Constrained creativity\xspace} 

\newcommand{\PP}[1]{\smallskip\noindent{\bf\textsc{#1}}\xspace}

\definecolor{kiink}{RGB}{0,105,85}
\newif\ifshowki\showkifalse
\newcommand{\ki}[1]{\ifshowki\par\noindent{\color{kiink}\textbf{[Key smarts]}~\textit{#1}}\par\nobreak\fi}

\newcommand{\PPP}[1]{
\noindent{\textit{\IfEndWith{#1}{.}{#1}{#1.}}}
}
\fi

\newcommand{\compactcaption}[1]{\vspace{-0.6em}\caption{\bf #1}\vspace{-1.1em}}

\newcounter{dplabel}
\newcounter{dpnmbr}
\renewcommand{\thedplabel}{\textbf{\thedpnmbr}}

\newcounter{packednmbr}
\newenvironment{packedenumerate}{\begin{list}{\thepackednmbr.}{\usecounter{packednmbr}\setlength{\itemsep}{0.5pt}\addtolength{\labelwidth}{-15pt}\setlength{\leftmargin}{\labelwidth}\setlength{\listparindent}{\parindent}\setlength{\parsep}{1pt}\setlength{\topsep}{0pt}}}{\end{list}}

\newenvironment{packeditemize}{\begin{list}{$\bullet$}{\setlength{\itemsep}{1pt}\addtolength{\labelwidth}{0pt}\setlength{\leftmargin}{\labelwidth}\setlength{\listparindent}{\parindent}\setlength{\parsep}{1pt}\setlength{\topsep}{0pt}}}{\end{list}}

\definecolor{newink}{RGB}{160,0,120}

\begin{document}

\date{}

\title{\Large \bf Agentic RCA for Internet-Scale Services Using Constrained Creativity}

\author{
{\rm Sayan Sinha\textsuperscript{1,3}, Vipul Harsh\textsuperscript{3}, B. Aditya Prakash\textsuperscript{1}, Vyas Sekar\textsuperscript{2,3}, Hui Zhang\textsuperscript{2,3}}\\
\textsuperscript{1}Georgia Tech \quad \textsuperscript{2}Carnegie Mellon University \quad \textsuperscript{3}Conviva
}

\maketitle

\begin{abstract}
System administrators of Internet-scale services need to resolve failure incidents to maintain  reliability of such services.  
Ideally, we want a troubleshooting system to be: (1) expressive to known and unknown incidents with high accuracy; (2) cost efficient at scale; (3) explainable to provide actionable insights operators can act on; and (4) entail low effort from the operators.  
 Unfortunately, most existing systems, including emerging LLM-assisted agentic workflows and structured frameworks for authoring 
  diverse RCA algorithms fall short of achieving all four requirements. 
 We present \sys, a novel agentic system for troubleshooting for Internet-scale services.  \sys embodies the paradigm of \textit{\structuredimagination} that combines the best of LLM-assisted automation and exploration with
  the explainability and efficiency of a structured  approach. Instead of allowing an LLM agent to write arbitrary code or generate arbitrary responses, we provide the agent a restricted DSL to generate its response via simple loop-free data flow programs. 
This DSL, equipped with high level operators for troubleshooting, makes \sys's output accurate, verifiable and explainable. 
On a mix of synthetic and real-world workloads, \sys achieves up to $62\%$ better accuracy compared to state-of-the-art solutions, while providing more explainable responses at up to $12\times$ reduced cost. 
\end{abstract}

\section{Introduction}
 \begin{figure}[t]
      \centering
      \includegraphics[width=0.8\linewidth]{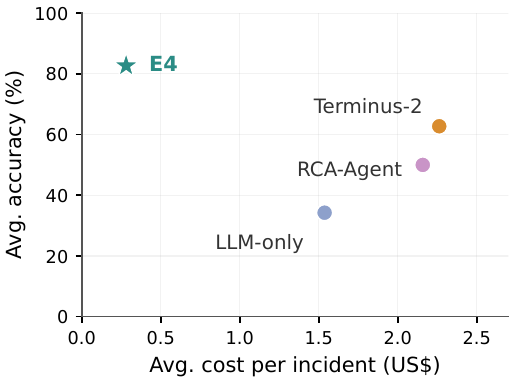}
      \compactcaption{\sys achieves higher diagnostic accuracy at lower cost than the evaluated agent-based RCA approaches across four datasets. Moreover, it produces more explainable analysis pipelines than the baselines.}
      \label{fig:accuracy-cost-summary}
  \end{figure}
Modern Internet services handle millions of clients in diverse and heterogeneous Internet conditions today.  Providers running such services care about key performance indicators or KPIs  (e.g., web page load time, latency, buffering) affecting user experience.
 To this end,  they constantly monitor and troubleshoot their systems to track and optimize these KPIs. Given the  dynamic operating environment and  heterogeneity of  clients, drops in KPIs are the norm, not the exception. When such KPI-impacting  incidents  occur, operators need   root cause analysis or troubleshooting workflows to  identify what caused the incident and initiate    mitigations.

While manually  navigating dashboards and issuing  queries may have sufficed in small deployments,   at  scale, automation of  incident diagnosis is critical.    In this regard, the community has responded with many solutions including specialized signature-based  and ML algorithms (e.g.,~\cite{murali2021scalable,sage,murphy}) and  high-level frameworks for   enable analysts to express their  hypothesis (e.g., ~\cite{pywhy_ecosystem,liu2023pyrca,moce2026}). More recently, there is significant interest in LLM-assisted agentic workflows  (e.g.,~\cite{xu2025openrca,shen2024openrca,rcaCopilot}). 

Ideally,  analysts    want  a root cause analysis system that is:
\begin{packeditemize}
    \item {\em expressive} to cover a wide range of known and unknown incidents with high accuracy.
    \item {\em efficient} in terms of the cost to process and diagnose incidents at scale.
    \item {\em explainable} so that they have actionable and understandable insights to act on the diagnosis; and
    \item entail {\em  low effort} to use. 
    \end{packeditemize}

Unfortunately,  existing solutions fail to simultaneously achieve  all of these   properties.  Classical signature-based systems and ML algorithms  can be  efficient and explainable. However, they lack the expressivity over  diverse and future incidents.  Using  a framework  to author flexible data analysis workflows can in   theory  be expressive and explainable. However, it requires significant  effort to design and  for  analysts to  learn to  write new capabilities.   LLM-assisted agentic systems have the potential to be expressive given the code generation and in-built world knowledge. 
Unfortunately, they  are also prone to inaccuracies, hallucinations, and unverified logic, and can  produce answers that are not explainable. Furthermore, LLM-based workflows  can  be inefficient in terms of the number of steps/tokens taken and data processing costs.

We present  \sys, an agentic system for root causing incidents in Internet-scale services that demonstrates it is possible to simultaneously achieve all four requirements. Our work builds on {\em \structuredimagination}~\cite{sinha2026constrained}, a recently proposed design paradigm for agents in systems operations. 
Instead of having an LLM generate arbitrary code or explore arbitrary hypotheses,  we constrain the creativity of the LLM to only produce programs in a  structured functional form using a specific DSL.
 Domain experts can provide an initial library of operators and diagnostic workflows expressed as dataflow DAGs  for  different kinds of incidents that may occur. The agentic system still has  the freedom  to reason when its library or playbooks are  insufficient and   dynamically  create new operators and DAGs.   Figure~\ref{fig:accuracy-cost-summary} previews our results:
  \sys combines higher average diagnostic accuracy with lower average
  cost than the evaluated baselines.

There are three  key  challenges  to translate this high-level idea into practice(\S~\ref{sec:newdesign}):  
(1) bootstrapping the operator library for a new domain; (2) resolving a new incident efficiently and explainably; and (3) evolving the operator library to tackle future issues. 
For (1), rather than use a generic domain- and deployment-agnostic library, we develop an effective heuristic to agentically adapt to the available telemetry types and data fields in a new deployment. 
For (2), rather than use a naive spray-and-pray approach and launching multiple parallel agents, we employ a  diagnostic approach that runs analysis via progressive stages of expressivity, moving to the next stage only if the previous stage was insufficient. 
Finally, for (3),  rather than naively allowing the agent to arbitrarily extend the operator library, we explicitly gate the admission of new operators based on their demonstrable utility to explain novel incidents. 

We develop a practical end-to-end prototype of \sys. We develop an initial  library of core operators and example playbooks for  common incident types.  We design a systematic methodology for hypothesis generation, pruning, and  scoring. 
We evaluate this prototype on a mix of real-world and synthetic incidents based on data from a large application  analytics  provider, including a small pilot study on real incidents. We find that \sys can be up to $62\%$ more accurate than the strongest baseline agentic system
 in terms of identifying the root causes within the top-3 suggested leads, and $5.4\times$ cheaper on average, and up to $12\times$ cheaper, in terms of total analysis cost. We run a number of factor analysis studies and find that the use of the abstraction in \structuredimagination is the key contributor. We show that \sys can be effective even with lower reasoning models. We also demonstrate that \sys is  flexible and that its effectiveness can be  amplified via analyst hints and predefined playbooks.

\section{Background and Motivation}

\begin{figure}
   \centering
   \includegraphics[width=\linewidth]{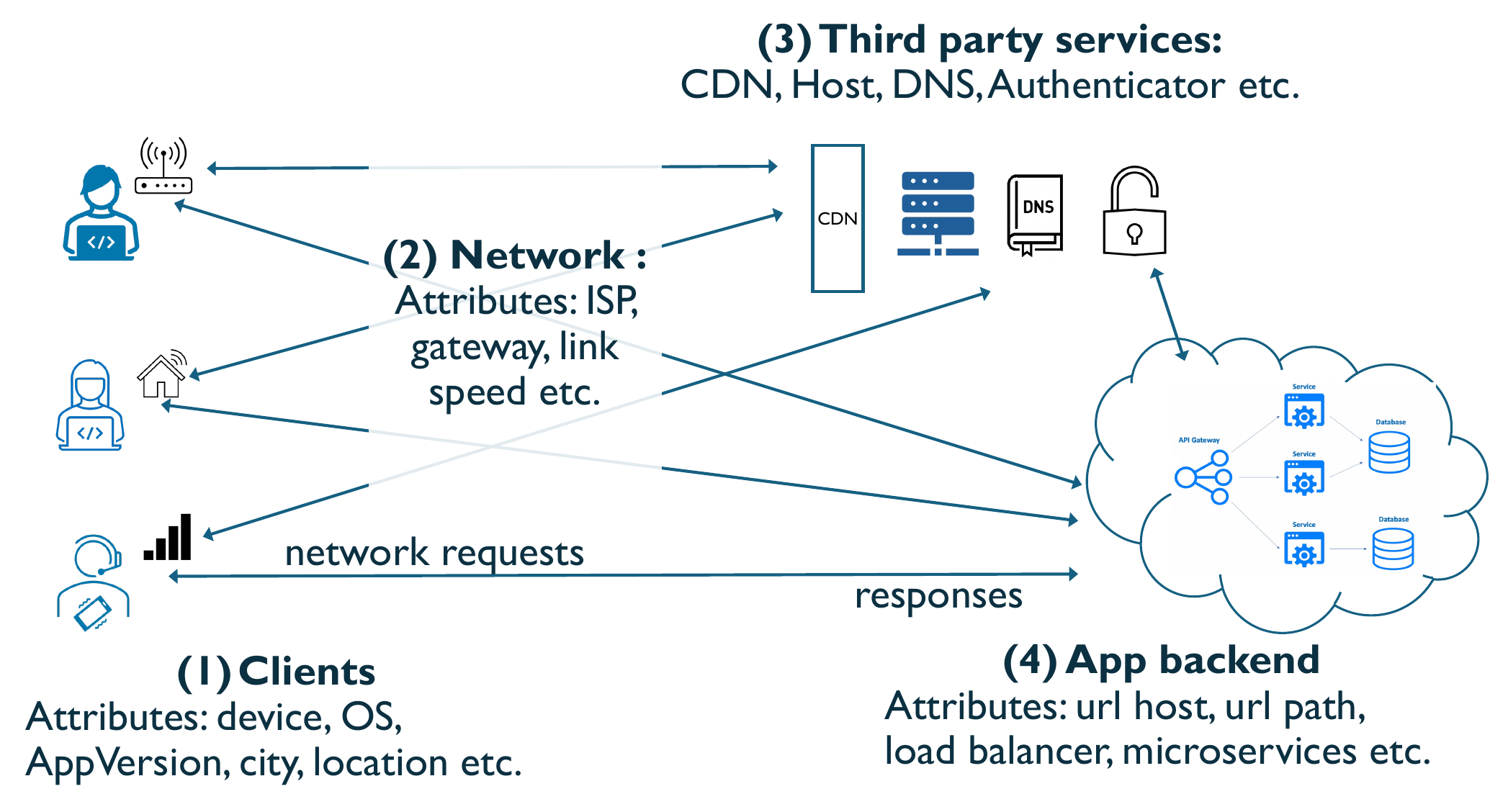}
   \vspace{-0.4cm}
   \compactcaption{Internet-scale services entail complex interactions among clients, network-services, third party services, and backend components and can fail in various ways impacting user-experience. }
   \label{fig:relationships}
\end{figure}

We begin with illustrative    troubleshooting and RCA scenarios  in large-scale services (Figure~\ref{fig:relationships}). We identify  key requirements that {\em any} such system must satisfy and argue why existing approaches fall short. 

\subsection{Motivating scenarios and requirements}

\PP{Setting} We consider an RCA assistant that runs alongside an existing alerting system. Given an alert that specifies a KPI anomaly (and its baseline/anomalous time windows), the assistant analyzes available telemetry data  and returns a rank-ordered set of hypotheses (e.g., a client cohort, backend service, user behavior sequence) with supporting evidence. The assistant can run fully autonomously from the alert alone, or accept an analyst hint that can prioritize a class of hypotheses and suggest how to probe them.

\PP{Illustrative scenarios}  We discuss three incidents inspired by true stories observed in a production service. We validated these using our simulation harness where we have more fine-grained control.

\begin{figure*}[t]
    \centering
    \begin{subfigure}[t]{0.32\textwidth}
        \centering
        \includegraphics[width=\linewidth]
        {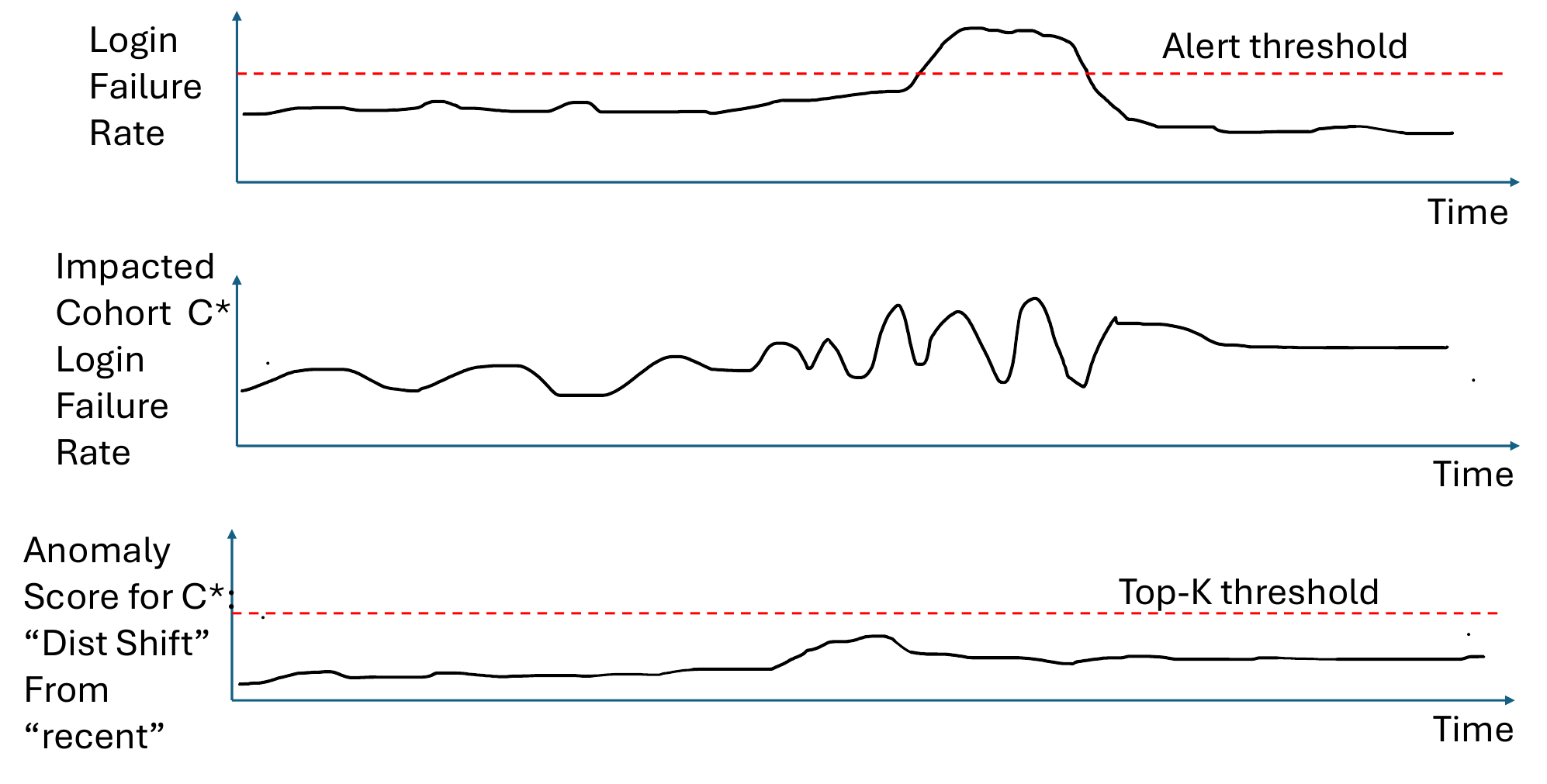}
        \caption{Example 1 (Login Failure): statistical test lacking second order or frequency-based analysis may miss the ground truth and produce false leads}
        \label{fig:motivation1}
    \end{subfigure}
    \hfill
    \begin{subfigure}[t]{0.32\textwidth}
        \centering
        \includegraphics[width=\linewidth]{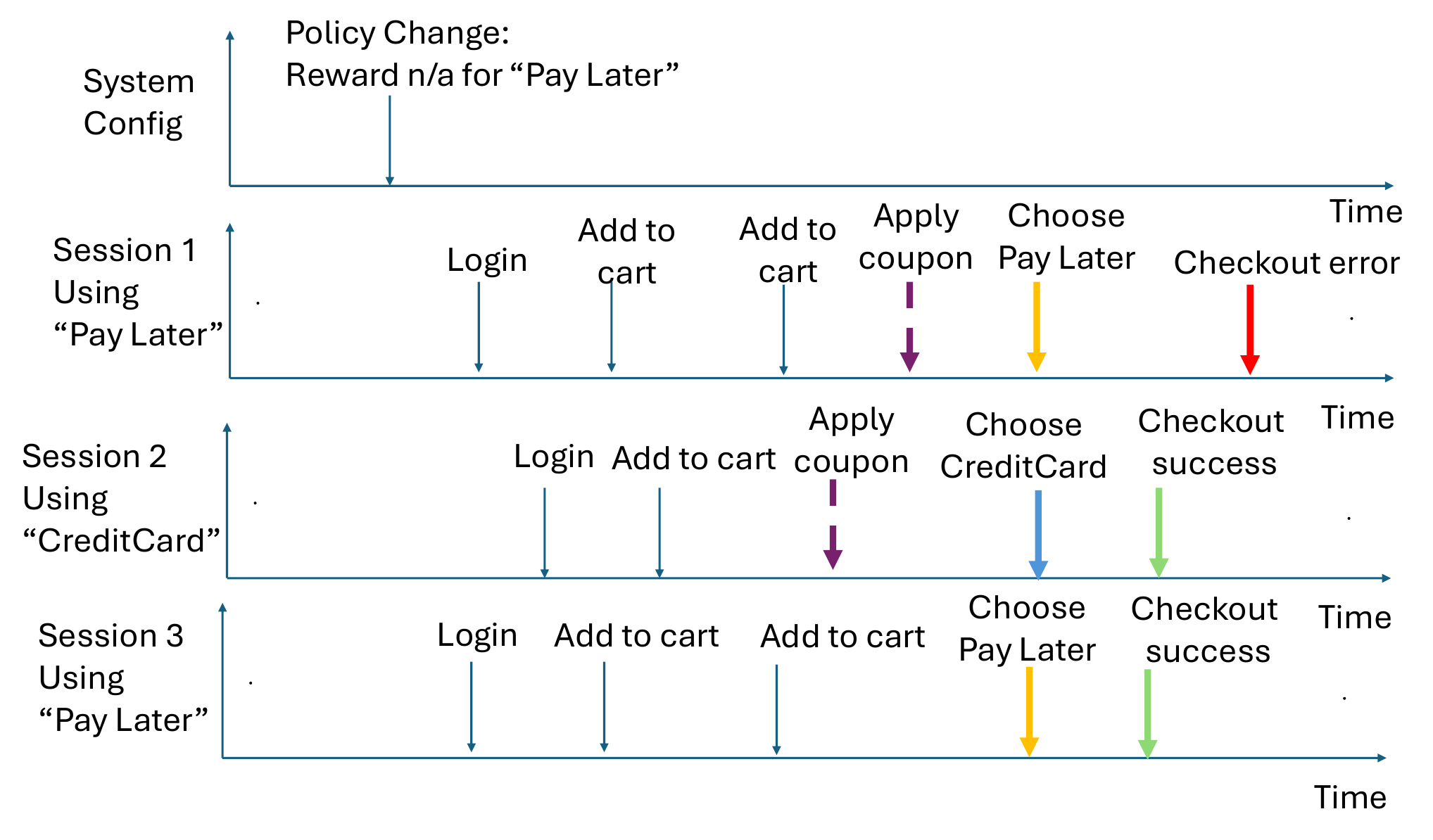}
        \caption{Example 2 (Checkout policy error): Looking at  static attributes misses  event sequences}
        \label{fig:motivation2}
    \end{subfigure}
    \hfill
    \begin{subfigure}[t]{0.32\textwidth}
        \centering
        \includegraphics[width=\linewidth]{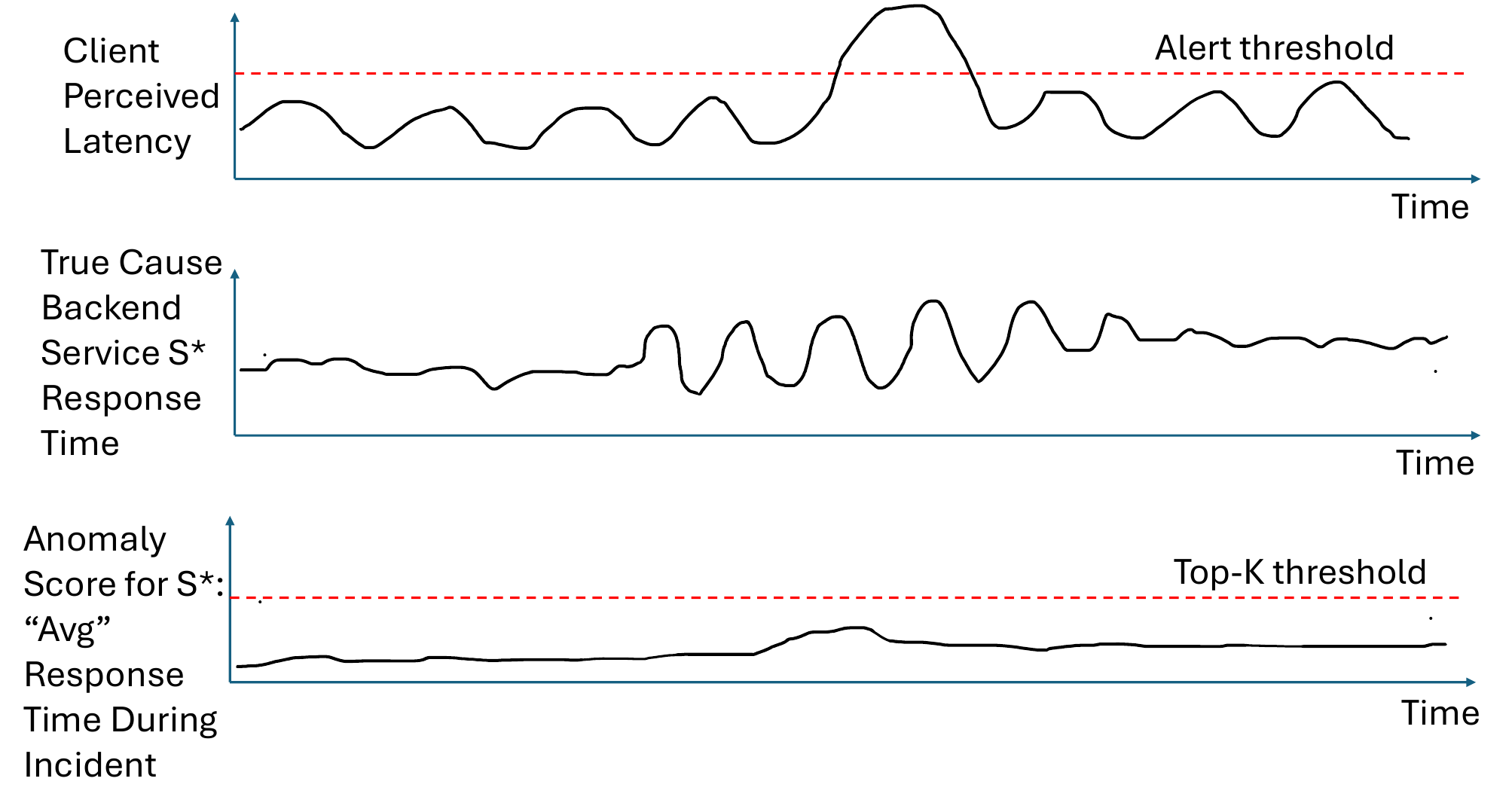}
        \caption{Example 3 (Latency Increase): Looking at just average values may miss  flapping phenomena}
        \label{fig:motivation3}
    \end{subfigure}
    \compactcaption{A simplified view of three real-world inspired incidents.  These examples illustrate that incidents can have diverse origins and patterns. They also show the   need  to  explore the space of possible hypothesis to explain  incidents.}
    \label{fig:motivation-ts}
\end{figure*}

\begin{packeditemize} 
    \item \textbf{Example 1 (Figure~\ref{fig:motivation1}):}
		    In this case, the KPI of interest to the provider is the login failure rate. 
            An alert fires that the login failure rate went up. 
            The root-cause of the problem is  that   clients belonging to a specific region were suffering higher failures.  The analyst used a traditional  distributional shift analysis. Here,     for each distinct cohort identified by client attribute values, they check  if  the distribution of login failures in recent times changed  significantly  and identify  top-5 cohorts that changed the most. 
            Unfortunately, this not only missed the true root cause but also produced  false leads. 
             In this case,  true problem cohort exhibited a change in the distribution with high variance much before the incident which eventually got amplified when the alert fired.  Ideally, the system should also  consider such {\em second order effects} on distribution properties or {\em frequency-based analysis}. 

            \item \textbf{Example 2 (Figure~\ref{fig:motivation2})} 
            In this example, the provider was trying to diagnose a sudden increase in the client checkout errors. But they could not find any discernible error or timeout  patterns in  backend logs or the client side logs. The true  cause turned out to be subtle. There was a policy update where  users using a specific  ``buy now pay later'' scheme were no longer allowed to use discount/coupon codes. Users who used this specific  payment option {\em after} applying the coupon code encountered checkout errors.  Other users using different payment modes or who did not  apply the coupon did not face issues. In this case, we need to look deeper into behavioral ``event sequence patterns'' (e.g., Coupon followed by PayLater)  that led to  errors.  
            
    \item \textbf{Example 3 (Figure~\ref{fig:motivation3}):}
            In a third  example, the provider was concerned about client-perceived latency. An alert fires on an end-to-end latency KPI. 
            The true root cause was that one of the backend services   handling  client requests was {\em flapping} and its service response time  caused clients assigned to it to suffer. 
             The analyst relied on a traditional ``correlation''  approach to look  for backend services whose average response time during the incident duration was high and identified the top-5 services. Unfortunately, these were all false leads and missed the true problem. %
           If the analyst  had some intuition that  flapping could explain it, they would like to dynamically craft a new hypothesis   focused on identifying  instability-focused tests that rank services by the change in coefficient of variation.
    
\end{packeditemize}

\PP{Requirements}
The above examples help us  identify four key  requirements   for {\em any} automated system for RCA in Internet-scale services: 
\begin{packeditemize} 
  \item Expressive: 
Our 
     examples show  the  need to consider a  diverse   set of  hypotheses that can  span client-side attributes, event patterns, backend metrics, and so on. They  also highlight that we may need to look beyond       rigid statistical tests or heuristics to avoid  blind spots and false leads.   

    \item Efficient:   At the same time, we cannot  do a brute force search over all possible hypothesis or comb through all data. Both cost and responsiveness of incident resolution are critical and we should be able to get to resolution or a set of leads quickly and cost effectively. 
         
    \item Explainable: The examples  involve non-trivial reasoning beyond basic heuristics and   second-order effects. In this case,    the analysis should be explainable, so that its outputs are actionable. Imagine a blackbox system that  states an answer or produces a long python notebook. Even if it was correct, the analyst would be uneasy to take the recommendation on face value. 
    
    \item  Effort: The system should be easy to use by practitioners, so they do not  deal  with complex  notebooks and writing Pandas/SQL queries.  Furthermore, human  analysts may want to see the output of AI algorithms and refine them based on their domain knowledge and steer the system or give it hints based on their domain knowledge (e.g., about service flapping in Example 3). 
   
\end{packeditemize}

\subsection{Related Work and Limitations} \label{rw1}
Troubleshooting and RCA are well studied  in academia and industry. 
At a high level, we can categorize  existing efforts into one of three classes as shown in Table~\ref{tab:rw} (with LLM-assisted agents having two subclasses). 

\begin{packeditemize}
\item {\em Specialized algorithms:}  Conventional RCA approaches use  custom algorithms for  different types of incidents in client-side and backend services (e.g.,~\cite{gan2021sage,harsh2023murphy,gestalt,shrink,sherlock,orion,netseer,gamma,errLog}).  While they  are mostly explainable  and efficient, they have two fundamental gaps in terms of expressivity and effort---limited in the types of incidents they can cover and effort to develop the algorithm.

\item {\em Structured frameworks for data analytics/RCA workflows:} In addition to bespoke algorithms for narrow incident classes, frameworks allow analysts to author new troubleshooting workflows (e.g.,~\cite{dowhy,liu2023pyrca,pywhy_ecosystem}). ExplainIt!~\cite{explainIT} provides a declarative SQL interface for exploring and ranking candidate causes in time-series data. MoCE~\cite{moce2026} expresses hypothesis exploration as operator DAGs and reuses computation across analyses. These frameworks provide flexible, structured analyses, but analysts still supply the queries, workflows, or new analysis methods needed to cover diverse incidents. \sys builds on this approach by using LLMs to compose analysis DAGs and synthesize new operators when needed.

\item {\em LLM-assisted RCA agents:} Recent efforts use LLMs to guide diagnosis and gather evidence through tools~\cite{xu2025openrca,wang2024rcagent,roy2024llmagents,hui2026chatrca,ye2026profrca}. Workflow-guided agents constrain investigations using predefined procedures~\cite{pei2025flowofaction,gao2026rcaflow,llexus,stepfly}, making the analysis easier to follow but limiting its expressivity. Coding agents can introduce new analyses on demand, but their generated code and reasoning trails can be difficult to inspect.
A second, and related issue is efficiency. The LLM may hallucinate and/or chase a number of dead ends, consuming time/tokens in the process~\cite{ai_fail_rca}. LLM-assisted agents also repeatedly examine telemetry and intermediate results, incurring substantial token costs. Finally, even with {\em tool augmentation} we observe that state-of-the-art (public) capabilities are quite limited in terms of effectiveness~\cite{wang2024rcagent,rcaeval,wang2026cloud,clark2026sregym}.

\end{packeditemize}

\begin{table}[t]\centering
\begin{scriptsize}
    \begin{tabular}{p{2cm}|>{\centering\arraybackslash}m{1.2cm}|>{\centering\arraybackslash}m{1.2cm}|>{\centering\arraybackslash}m{1cm}|>{\centering\arraybackslash}m{1cm}}
        System Class  &Expressive &Efficient &Explainable & Effort  \\ \hline  
        Specialized algorithms  (e.g., \sage \cite{gan2021sage}, \murphy \cite{harsh2023murphy}, \cite{gestalt,shrink,sherlock,orion,netseer}) &\xmark &\cmark &\cmark &\xmark \\ \hline

Frameworks for authoring data analysis workflows    (e.g., SQL, ExplainIt!~\cite{explainIT}, \moce~\cite{moce2026}) &\xmark &\cmark &\cmark &\xmark \\ \hline
        
        Workflow-guided RCA agents (e.g., LLexus~\cite{llexus}, StepFly~\cite{stepfly}) &\xmark &\xmark &\cmark &\cmark \\ \hline

        Coding agents for RCA (e.g., \terminus~\cite{harbor2026}, RCA-Agent~\cite{xu2025openrca}) &\cmark &\xmark &\xmark &\cmark \\ \hline

        \textbf{Ours (\sys)} &\textbf{\cmark} &\textbf{\cmark} &\textbf{\cmark} &\textbf{\cmark}  
    \end{tabular}
    \end{scriptsize}
    \caption{Qualitative comparison against the four requirements of modern RCA. Ticks indicate support for the stated requirements; crosses indicate limitations.}\label{tab:rw}
    \vspace{-0.6cm}
\end{table}

\medskip 
\noindent  {\bf Takeaways:}  We see that none of the current paradigms of
RCA/troubleshooting efforts  can simultaneously achieve all of our requirements.
Each class achieves only a subset of these requirements at the expense of the
others; an ideal solution would satisfy all four simultaneously.
Next, we discuss how we can indeed achieve this in practice.

\begin{figure}[t]
    \centering  
    \centering
    \includegraphics[width=0.87\linewidth]{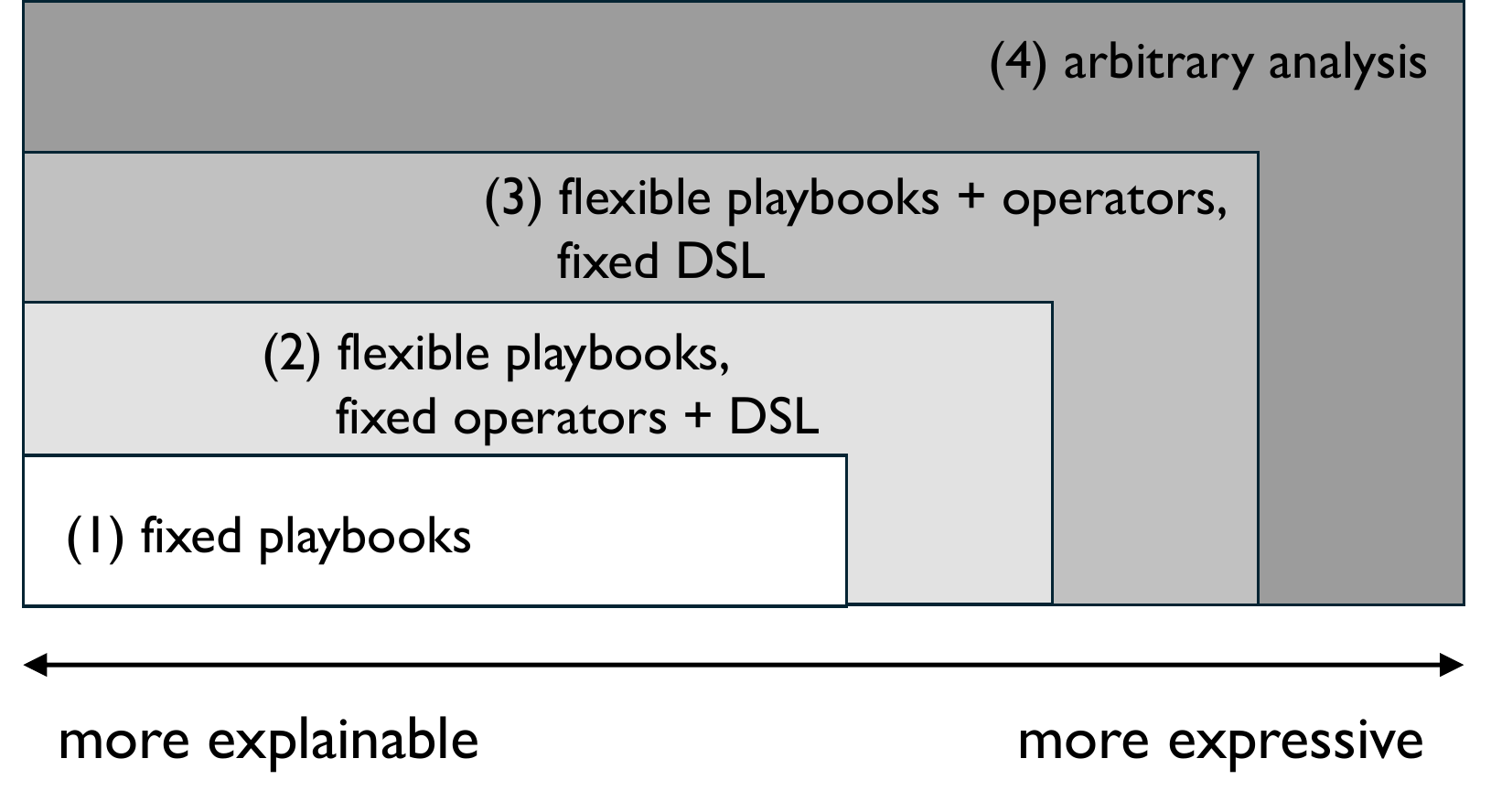}
    \compactcaption{The inherent tradeoff between explainability and expressivity among RCA approaches: 
    (1) RCA is performed via fixed pre-defined playbooks, 
    (2) RCA is performed via composing new playbooks based on existing DSL and operators,
    (3) RCA is performed via authoring new operators in an existing DSL and then composing new playbooks using those operators,
    (4) the RCA agent can write arbitrary code to orchestrate the analysis. \sys confines RCA to (1), (2) and (3), using more expressive power only if the previous level was insufficient. 
    }
    \label{fig:dsl_venn}
    \label{fig:design-space}
    \label{fig:spectrum}
\end{figure}

\section{Design Overview} \label{sec:overview}
\ki{We have identified that existing agentic RCA systems make a tradeoff between expressivity and explainability}

\subsection{High-level Idea}
Every analysis  in an RCA system can ultimately be thought of as a data analysis program. The program takes the telemetry data, associated metadata, and the anomalous KPI as input, and outputs a ranked list of root cause leads.
We posit that high expressivity, explainability, and efficiency are desirable traits of this program, and that an RCA system should produce it with low effort.

To achieve low-effort, we 
adopt an LLM-driven RCA system similar to prior work.  That said, there is a wide spectrum of how we can use an LLM. %
Different approaches  in  agentic RCA systems have different levels of expressivity and explainability on a  \eespectrum as shown in   Figure~\ref{fig:spectrum}.
At the expressivity end are general-purpose coding agents, which can write any program. Close to this end, RCAgent~\cite{wang2024rcagent} and the agent framework evaluated in OpenRCA~\cite{xu2025openrca} enable flexible tool use and code generation, making their programs expressive but hard to explain. Moving toward greater explainability, works like Flow-of-Action~\cite{pei2025flowofaction} follow previously created SOP-guided investigation with some procedure generation. At the explainability end are works that only execute human-authored troubleshooting guides, such as LLexus~\cite{llexus} and StepFly~\cite{stepfly}, where the procedure is fixed in advance and the agent's role is limited to carrying out its steps.

\ki{\structuredimagination describes the region in the spectrum (what) that needs to be explored and how to explore it}
\ki{We propose the region in the spectrum where an RCA system should lie, instead of picking one particular spot.}

To navigate this tradeoff, we adopt \textbf{\structuredimagination}~\cite{sinha2026constrained}, a paradigm in which the programs an agent produces stay within a bounded region of this \eespectrum.  More specifically, we {\em constrain} the structure that  every analysis program must follow using a specific domain-specific language (DSL).   As illustrated in \cref{fig:dsl_venn}, the DSL and operator library  restricts the space of programs that can be generated to perform the analysis.    In essence, we can think of each RCA analysis  as a {\em composition} of some basic lego blocks or {\em operators} that retrieve or analyze the telemetry data, prune the space of related entities,  construct the trajectories, and perform some  statistical or contrastive  analysis to produce leads. 
Said differently, our DSL defines a vocabulary of callable \emph{operator} APIs, such as generating candidate hypotheses or scoring them, and a grammar requiring that calls to these APIs compose into an analysis DAG. The grammar is fixed, so every analysis program obeys the same structural rules. This structure also helps keep token and execution costs low (\S\ref{sec:runtime}).

Realizing this paradigm raises four questions: how to represent the structure that every analysis program must follow; what vocabulary that structure can use, and where it comes from; how to keep the cost of an investigation in check; and what to do when the vocabulary does not suffice. We answer these in the following  sections.  

\begin{figure}[t]
    \centering
        \includegraphics[width=1.05\linewidth]{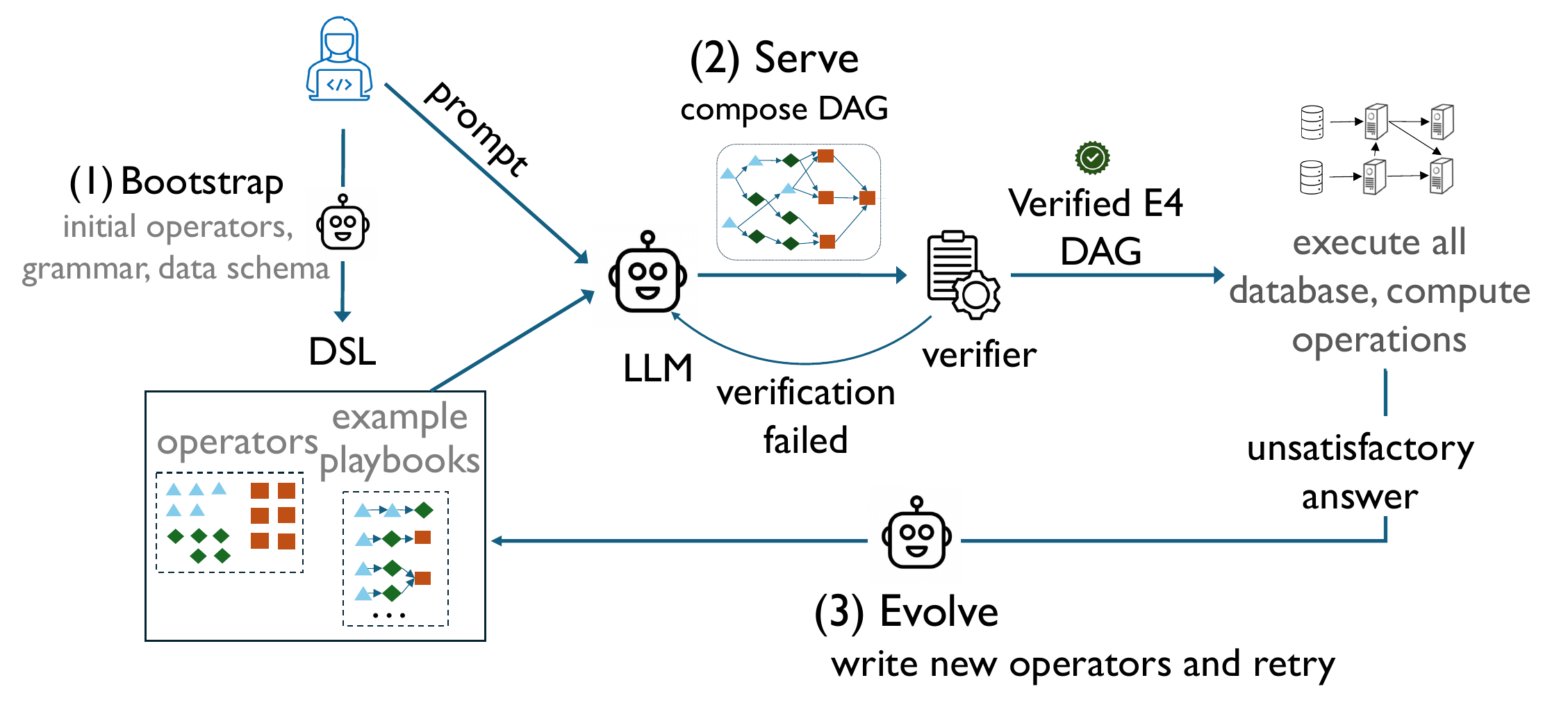}
    \compactcaption{\sys is a practical realization of   \structuredimagination with 3 high-level ideas: (1) Bootstrap: adapting the DSL operators to the specific domain, incorporating any domain knowledge or custom data columns, (2) serving an incident by crafting a DSL-compliant DAG for the diagnosis and (3) if the current operator library is not sufficient, \sys authors new operators and reruns diagnosis. 
    }
    \label{fig:system-workflow}
\end{figure}

\PP{Relation to harness engineering} Skills and harness engineering decide what the agent can call, how calls compose, and what executes. Harness design has grown from prompt templates into runtime and systems design~\cite{weng2026harness}, spanning workflows~\cite{hong2024metagpt}, evaluation, permission control, state management, and review loops~\cite{shinn2023reflexion}. For RCA, \structuredimagination can serve as the basis for designing a good harness based on its DSL in conjunction with these techniques.

\subsection{System Overview And Challenges}
We present   \sys,   a practical realization of this  high-level idea of   \structuredimagination. To fix the  grammar  every analysis program must follow,  we borrow the DAG structure of a recent work~\cite{moce2026}. \sys ships with an initial set of operators common to RCA in Internet-scale services; this vocabulary can be extended during the investigation of an incident.   \sys follows a two-step exploration process of the given region of \eespectrum. At level 1 -- the DSL has a given vocabulary and grammar and \sys is restricted to use the given vocabulary and grammar to explore the space of all possible analysis programs and come up with the best that solves the anlysis. At level 2, the vocabulary can be evolved using an LLM, thereby increasing the space of programs more towards the space of all programs supported by the DSL's grammar. These two steps are analogous to the two levels in \cref{fig:spectrum}. 

Figure~\ref{fig:system-workflow} gives an end-to-end view of \sys. Panel (a) shows the DSL: operators with compatible input and output types, composed into an analysis DAG. Panel (b) shows the three-stage recipe for building, running, and evolving the agent. An initial operator library and example playbooks provide the starting vocabulary; the recipe grows the vocabulary while the grammar stays fixed. The recipe is agnostic to the choice of LLM and comprises the following stages:

\begin{packedenumerate}
    \item \textbf{Bootstrap} runs once per deployment and fixes the vocabulary the agent starts from, and with it the program space of \cref{fig:dsl_venn} that level~1 explores. The LLM takes as input the initial operators and example playbooks, the deployment's telemetry schema, a description of the RCA task, and optional analyst hints. From these it generates additional operators from the schema and domain knowledge of the task, then composes example playbooks over them. New operators must use input and output types supported by the DSL; the grammar remains fixed. The LLM sees the schema but never the deployment's telemetry. The playbooks undergo static verification and execution on synthetic data generated from the schema to catch runtime errors. Bootstrap hands the agent a deployment-specific operator library and verified playbooks for use at runtime.

    \item \textbf{Runtime adaptation} runs per incident, at level~1 of \cref{fig:spectrum}. The agent is given the incident as a natural-language description, relevant metadata, and optional troubleshooting hints, along with the operator library and playbooks from bootstrap and access to the telemetry. It selects a playbook or composes a new DAG from the operators. The verifier checks that the referenced operators exist, connected operators have compatible types, and the graph is acyclic. Once verified, the DAG is passed to the execution engine, which schedules its database and compute operations. The agent can revise the DAG over several iterations to refine the diagnosis, and returns a ranked list of root cause leads; if no program over the current vocabulary yields a satisfactory result, it escalates to evolution.

    \item \textbf{Evolution} runs when runtime adaptation escalates, moving the agent to level~2 of \cref{fig:spectrum} and enlarging the space of programs it can compose (\cref{fig:dsl_venn}). The LLM is given the escalated incident together with the history of failed attempts, the current operator library and playbooks, and the task description. It proposes new operators, with input and output types supported by the DSL, that fill the gap the failures reveal, and composes playbooks over them; the grammar remains fixed. The extended library is admitted only if it lets the escalated incident be served, in which case it replaces the current one and is reused in subsequent investigations; otherwise the library is left unchanged.
\end{packedenumerate}

\section{Detailed design} \label{sec:newdesign}

This section provides detailed designs of the bootstrap, compose and evolve phases of our RCA agent, \sys, for a given deployment. Figure~\ref{fig:system-workflow} outlines the various components of the end-to-end system. There are 3 core ideas in the design: (1) bootstrapping \sys for a new domain, (2) diagnosing an incident via successive stages of expressivity and (3) evolving the library of operators in the DSL if the current operators are insufficient. We describe these next. 

\subsection{Bootstrapping \sys for a new domain} \label{sec:bootstrap}

Different internet services have different sets of available monitoring data with varying data schemas. The bootstrapping phase, executed once per deployment, 
adapts \sys to the specific system, 
ensuring the DSL includes obvious operators to incorporate domain-knowledge or take advantage of domain-specific columns in the data schema. For instance, in one scenario, the bootstrapping phase produced an operator that uses parent-child relationships in traces to rank services by their failed spans with no failed child, helping distinguish potential error origins from propagated failures.

For bootstrapping, we start with a starter DSL with an initial set of operators that are general for different modes of telemetry (e.g., timeseries, events, traces). Then we prompt a state-of-the-art coding agent with the following inputs: the initial operator set, the deployment's telemetry schema and
metadata, a natural-language description of the domain and the RCA task, a few example playbooks (pre-defined programs in the DSL) and an optional analyst hint. Based on this input, the agent synthesizes a specialized operator library and a playbook library customized to that deployment. %

The \textit{grammar} of the DSL constitutes the structure that \sys uses to compose the operators.  %
We fix \sys's grammar to a DAG of operators, but we note that other structures like tree, or a program with loops are also possible. We made this choice as previous RCA algorithms fit naturally into a DAG~\cite{moce2026}\footnote{refer to appendix~\ref{app:dsl-grammar} for the exact DSL grammar}. We next describe the starter DSL \textit{vocabulary}: the initial set of operators provided as input to the bootstrapping phase.

\begin{itemize}[leftmargin=*]
    \item \textit{Candidate generation operators}: Operators to analyze (1) data cohorts based on column attributes e.g., region, deviceVersion, (2) event sequences e.g., error sequences in the client-side device, (3) performance metrics e.g., CPU usage in backend services. They output a list of candidate root cause leads.
    
    \item \textit{Candidate scoring operators}: These include (1) statistical scorers that test a candidate’s metrics against the baseline (e.g., from history or another user cohort) via standard statistical test, (2) Bayesian models, such as decision tree, for learning distribution of metric values in fine-grained data cohorts and performing MLE. They output candidate root cause leads along with a score.
    \item \textit{Summarizer operators}: These include (1) ensemble aggregation operator to combine results from multiple algorithms via weighted majority, giving weight to root causes flagged by multiple algorithms (2) weight readjustment operators to distinguish primary root causes from secondary symptoms e.g., by penalizing the scores of upstream backend services that get affected by a failure. They output a ranked list of candidate root-cause leads.
\end{itemize}

If the analysis uses an ensemble of algorithms, the DAG-based DSL naturally avoids duplicate computation across different algorithms by utilizing the same intermediate result for different downstream analysis.

\sys's operators abstract away implementation complexity from the agent; it only needs to know the functional signature describing the input/output types of each operator.
For instance, the candidate generation operators require issuing complex SQL queries (e.g. to get all events after login-start and before login-end for each user-session) or perform complex stateful computation over the event stream (e.g. to compute if a user is a frequent user or not). Similarly, the scoring operators have complex algorithmic machinery e.g., employing Aho-Corasick data structure to handle ``event strings'' for
finding event patterns. The agent does not need to know or handle any of these complexities, which boosts its serving accuracy and context usage. These operators are also optimized for large-scale data, resulting in better efficiency. 

\subsection{Serving incidents via progressive stages}\label{sec:composition}\label{sec:runtime}

Once the DSL operator library gets updated via bootstrapping, it can be used for diagnosing incidents. \sys's progresses in stages with each stage providing more expressivity if the analysis from the previous stage was not satisfactory (see illustration of stages in Figure~\ref{fig:dsl_venn}). The rationale here is constraining the creativity of the agent makes the responses more explainable, so we only employ more expressivity if needed. 

To troubleshoot an ongoing incident, a human prompts the \sys agent, optionally with any domain knowledge about the failure. The operator descriptions and the example playbooks are passed along with the prompt. The \sys agent then composes a DAG for RCA, in successive stages (Figure~\ref{fig:dsl_venn}). We describe the first two stages below:

\begin{itemize}[leftmargin=*]
    \item \textbf{Stage (1) Fixed playbook}: the \sys agent picks an existing playbook from the set of given playbooks for performing RCA and instantiates its hyperparameters for the given incident. For instance, the agent can select a playbook to analyze client cohorts and look for corresponding root causes e.g., browser version, operating system, backend service. If the root cause is not among these entities or it is not found using the chosen playbook, the agent moves to the next stage.
    \item \textbf{Stage (2) Flexible playbook, fixed operators}: the agent composes a new DSL-conformant DAG using the existing operator library. The agent can pick the hyperparameters for individual operators in the DAG. For reference, Figure~\ref{fig:market-incident-dag} in appendix includes a detailed DAG produced by \sys.
\end{itemize}

Once the agent generates a DAG, \sys's verifier ensures that ``the DAG compiles'' before passing the DAG to the runtime execution engine. We describe the verifier and the runtime execution engine briefly before moving on to the next stage of evolution (stage 3 in Figure~\ref{fig:dsl_venn}).

\paragraph{Verifier}
We employ a simple verifier to ensure the candidate DSL DAG generated by the \sys agent is semantically correct, thereby increasing the reliability of the compose phase. Specifically, the verifier checks for the following: 

\begin{itemize}[noitemsep,topsep=1pt,leftmargin=*]
    \item operator validity to prevent hallucinated operators
    \item type-safety to ensure input/output data types match for each operator in the DAG
    \item the operators are composed via a DAG and do not include any cyclic calls
    \item candidate generation operators are consumed only by scoring operators or by other candidate generation operators, and no operator remains unused
    \item the program finally ends in exactly one answer
\end{itemize}

\paragraph{\sys's Runtime Execution Engine}

The runtime system is responsible for scheduling all the database and compute operations of the generated DAG, parallelizing them across a distributed cluster. Note that the DAG operations can be of different types e.g., SQL, ML inference, stateful filtering. The runtime also caches intermediate computation to save execution costs, indexing a DAG node by computing a recursive hash of its resolved inputs. This cached result can be resued if the agent invokes the same subgraph in a later analysis. It can also be used for failure recovery in case one of the parallel nodes fails mid-execution. \sys allows the user to change the maximum size of this cache along with the cache eviction policy e.g., Least recently used or Last in first out. 

Note that at no stage does the LLM  interact with data directly. The agent  reads the prompt, the operator descriptions and the playbook library and produces a DAG. The DAG gets executed in a local environment without any agent interactions. This has two advantages. Firstly, this results in low token costs without sacrificing the accuracy of the analysis (\S~\ref{sec:end-to-end}). Secondly, this naturally addresses sovereignty and privacy , if any, for sensitive and proprietary data as no data needs to leave the enterprise. 

\subsection{Evolving \sys's operators for more expressivity} \label{sec:evolution}

When the second stage (flexible playbook, fixed operators) of the agent also fails, it implies no program over the current vocabulary (operator library) could diagnose the incident, and the operator library must be extended. This represents \textbf{stage (3)} of expressivity (flexible playbook, flexible operators, fixed DSL) in Figure~\ref{fig:dsl_venn}. This step can be thought of as a joint optimization problem (Appendix~\ref{app:objective}) to generate the most accurate program (P) from \sys's operator library (L) and both P and L are modified alternatively. 
For synthesizing operators, \sys analyzes diagnosis gaps. It retains new operators only when they improve the diagnosis, avoiding ``operator-bloat''.

To evolve the operator library, we take the escalated incident together with the results of every program tried and why it was deemed insufficient and pass them as input to a coding agent. The LLM is prompted to first evaluate the \emph{gap}, \ie\ why previous attempts
were insufficient, and then to propose new operators based on this evaluation. The serve phase of \sys's agent is then invoked again, to diagnose the incident using the new operators. Ultimately, the new operators are admitted to the DSL permanently, if they produced the true root cause (decided via user feedback).  

As an example, in one incident, the gap evaluation showed that no operator in the library generates root cause candidates corresponding to the endpoint of a request, and asks for suitable operators. The evolve phase, in this case, synthesized two new operators: 
the first operator emits candidates based on requests' target endpoint (host and path), filtering out rarely used endpoints. The second operator computes the 
candidate's failure rate within a window by a count-weighted mean and standard deviation and scores the candidate by the Wasserstein-2 distance between the baseline and anomalous windows. 
Both were 
admitted, taking the library from $19$ operators to $21$. \sys's agent then orchestrates new analysis by using these two operators along with existing cohort-based summarizer operators. This updated analysis  produced the right root cause as the top lead: clients in a single geographic region were experiencing high login-failure rates ($74\%$).

\section{Implementation} \label{sec:implementation}

\paragraph{System and runtime setup} \label{sec:execution-engine}
We implement \sys in about 7.5K lines of Python and run the prototype on a
single MacBook Pro with an Apple M3 Pro CPU (12 cores: 6 performance and
6 efficiency), 18\,GB of unified memory, and macOS. Programs are serialized as
JSON: each node names an operator and its inputs, with node identifiers
representing dependencies. The execution engine uses \texttt{Ray}
tasks/actors~\cite{moritz2018ray} to run independent branches in parallel and
caches node outputs for reuse across attempts. Operator admission checks
Python syntax, declared signatures and return annotations, and scans the AST
for prohibited I/O and dynamic-code constructs; colliding operator names
receive new versions. We run Python 3.10 and \texttt{Ray} 2.49. Across all experiments, the LLM
is Gemini-3.1-Pro, served remotely through Google Vertex AI; only the operators
execute locally.

\paragraph{Configuration}
The runtime uses binary analyst feedback: it accepts a diagnosis
when the analyst is satisfied and otherwise proceeds to the next
stage, subject to the attempt budget.\footnote{Table~\ref{tab:params}
in Appendix~\ref{app:settings} lists the default runtime,
LLM, and data-processing settings.}

The digest reports baseline and anomalous aggregates side by side: row
counts, per-entity failure rates, and latency quantiles $p_{50}/p_{90}/p_{99}$.
The data-processing limits cap the largest entity groups, time
buckets and string lengths. Bootstrap tests playbooks on schema-derived
synthetic telemetry covering five fault types: cohort failure, cohort
duration, latency, CPU stress and primary URL failure. Generation and
scoring operators use separate synthesis prompts. Library eviction removes the
least-used operators, counting usage over incidents seen so far, and drops
playbooks that no longer verify.

Operators do not carry fixed statistical thresholds. Every threshold, such as
a significance level, a minimum support or a minimum traffic share, is a
declared parameter with a default and an optional estimator; a playbook wires
it as \texttt{auto} and the engine estimates it from the baseline window at
execution time. $\mathrm{OVer}$ rejects an operator that hard-codes a
threshold in its body, so a value the agent cannot see or set never decides an
answer.
\begin{table}[t]
\centering
\scriptsize
\setlength{\tabcolsep}{3pt}
\begin{tabular}{@{}p{0.29\columnwidth}p{\dimexpr0.71\columnwidth-6pt\relax}@{}}
\toprule
\textbf{Family} & \textbf{Inputs and computation} \\
\midrule
\multicolumn{2}{@{}l}{\emph{\textbf{Data access}}} \\
Loaders and aggregators & Read sessions and aggregate telemetry by window and attribute. \\
\midrule
\multicolumn{2}{@{}l}{\emph{\textbf{Candidate generation}}} \\
Client cohorts & Propose cohorts by region, ISP, OS or device version. \\
Backend metrics & Propose services with CPU, memory or disk conditions. \\
Event patterns & Derive candidate subsequences from session events. \\
URL services & Propose backend or third-party URLs with failures or delays. \\
\midrule
\multicolumn{2}{@{}l}{\emph{\textbf{Scoring}}} \\
Differential diagnosis & Compare cohort, URL or server metrics against baseline using t-tests or permutation tests. \\
MLE & Fit a decision-tree regressor to historical cohort KPIs; score cohorts by how well they explain the anomaly. \\
Flow model & Fit a weighted sum of URL metrics to the KPI; score changes in each URL's contribution. \\
Event pattern & Use Aho--Corasick matching to compare subsequence support in failed and successful sessions. \\
\midrule
\multicolumn{2}{@{}l}{\emph{\textbf{Summarization}}} \\
Weight readjustment & Use backend dependencies to downweight propagated symptoms and boost likely primary causes. \\
Ensemble aggregation & Combine scores, giving more weight to candidates supported by multiple scorers. \\
\bottomrule
\end{tabular}
\compactcaption{Starter operator families and their computations.}
\label{tab:operator families}
\end{table}

\paragraph{Default operators and playbooks} \label{sec:shipped-operators}
The starter library contains 19 operators: 7 candidate hypotheses generators, 10 scorers, and 2 summarizers. Eleven example playbooks ship with it. Table~\ref{tab:operator families} summarizes the operator
families grouped by their function. The playbooks cover client cohorts,
backend services, URL-driven incidents, and combined client/backend analysis.

\section{Evaluation}

We evaluate \sys on four public benchmarks and on
a custom  benchmark.
We first compare overall accuracy and cost across these datasets, then
examine incident types, recipe stages, and sensitivity to workload conditions.
 
\subsection{Setup} \label{sec:eval-setup}
\PP{Datasets}
We evaluate on the \market, \bank, and \telecom deployments from
\openrca~\cite{xu2025openrca} and ORCA-bench~\cite{gong2026orca}, which
provide backend telemetry. To additionally evaluate client-side diagnosis,
we extend DeathStarBench's Social Network application~\cite{gan2019open}
with simulated clients, attributes, and session-event histories.
Table~\ref{tab:datasets} summarizes the datasets; Appendix~\ref{app:dataset-details}
describes their telemetry, KPIs, faults, and preparation.
The DSB corpus contains 64 incidents. For ORCA-bench, we construct
windows from frontend KPI anomalies without using fault-injection times.

\begin{figure*}[!t]
    \centering
    \begin{subfigure}{0.97\linewidth}
        \centering
        \includegraphics[width=\linewidth]{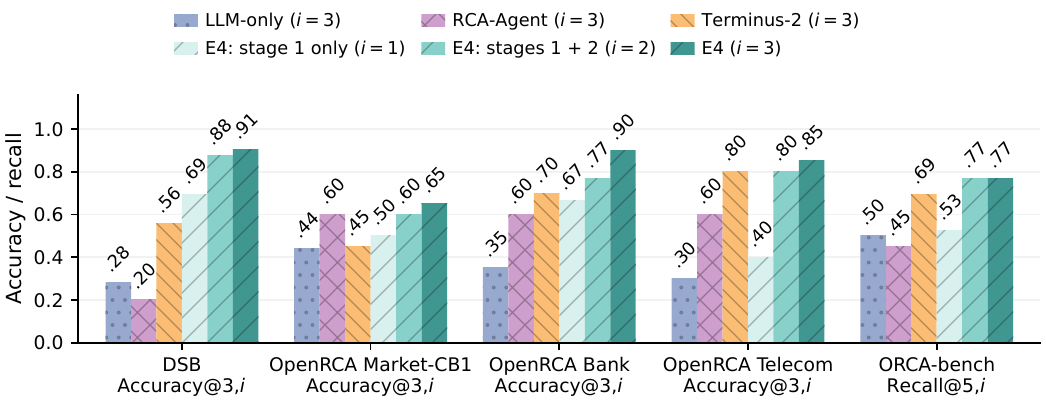}
        \caption{\sys achieves 65-91\% accuracy across all benchmarks compared to $\leq 56\%$ for baselines. The accuracy metric is indicated for each dataset and the attempt budget $i$ is shown in the legend. Baselines use up to three attempts; each \sys stage adds one attempt.}
        \label{fig:market}
    \end{subfigure}
    \\[0.12cm]
    \begin{subfigure}{0.97\linewidth}
        \centering
        \includegraphics[width=\linewidth]{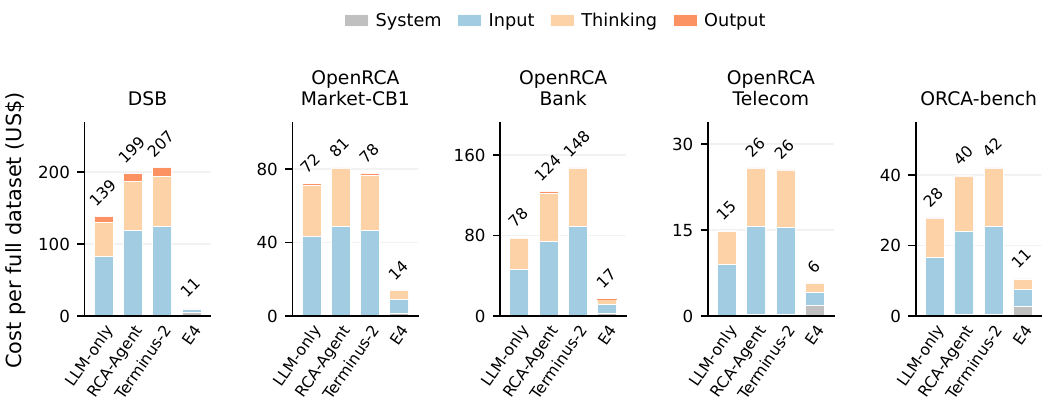}
        \caption{Full-dataset cost in US\$ by component (lower is better).}
        \label{fig:market-cost}
    \end{subfigure}
    \compactcaption{Overall results on the five datasets. Compared with \terminus, \sys achieves $1.3\times$ the accuracy or recall and reduces cost by a factor of $8.1$ on average, using the mean of per-dataset ratios.}
    \label{fig:openrca}
\end{figure*}

\begin{table}[t]
\centering
\footnotesize
\setlength{\tabcolsep}{3pt}
\begin{tabular}{@{}lp{0.34\columnwidth}rr@{}}
\toprule
\textbf{Dataset} & \textbf{Application / scope} & \textbf{Incidents} & \textbf{MB/incident} \\
\midrule
DSB & \raggedright Social network; client and backend & 64 & $479 \pm 251$ \\
\market & \raggedright Online shopping; backend & 70 & $224 \pm 37$ \\
\bank & \raggedright Banking; backend & 136 & $83 \pm 56$ \\
\telecom & \raggedright Telecommunications; backend & 51 & $107 \pm 3$ \\
ORCA-bench & \raggedright Online shopping; multiple backend causes & 36 & $221 \pm 50$ \\
\bottomrule
\end{tabular}
\compactcaption{Dataset summary. Distinct incident counts cover the full benchmark. MB/incident reports mean $\pm$ standard
deviation of compact JSON inputs in our evaluation.}
\label{tab:datasets}
\end{table}

\begin{figure*}[!tp]
    \centering
    \includegraphics[width=0.95\linewidth]{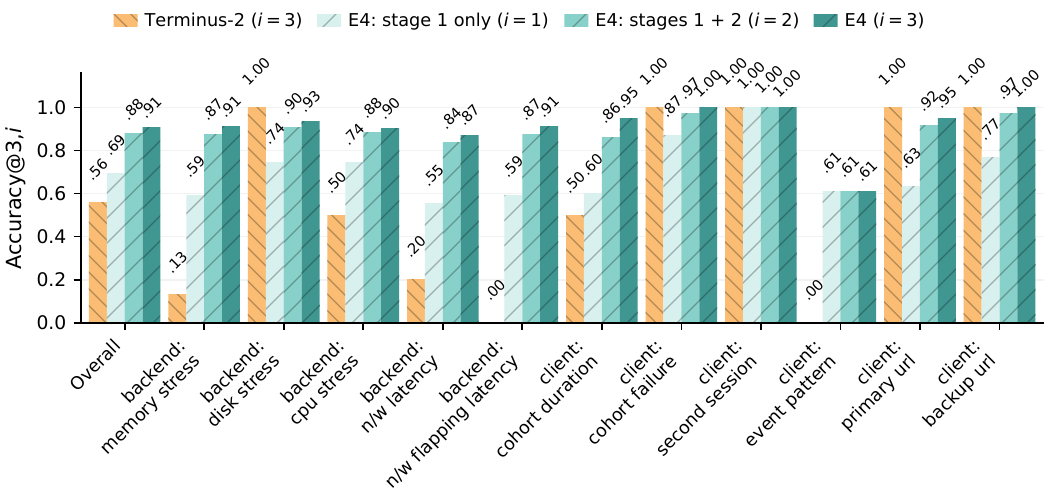}
    \vskip -0.4cm
    \caption{Accuracy@3,$i$ on DSB by incident type, with attempt budget $i$ shown in the legend. \sys has an average gain of $33.7$ percentage points over \terminus across the 11 incident types.}
    \label{fig:acc}
\end{figure*}

\PP{Baselines}
We compare \sys with three baselines. \emph{LLM-only} directly analyzes raw
telemetry selected using the balanced sampling approach in
\openrca~\cite{xu2025openrca}. \emph{RCA-Agent}, also introduced in
\openrca~\cite{xu2025openrca}, uses two LLM agents: a controller directs the
investigation and interprets results, while an executor writes and runs Python
code in a persistent Python interpreter. \terminus, the agent harness used in the ORCA-bench
evaluation~\cite{gong2026orca}, investigates raw telemetry through an
interactive Bash shell. We run all three baselines through the Harbor
harness~\cite{harbor2026} used by \terminus.
All baselines receive the same anomaly trigger as \sys,
including the anomalous KPI and historical/anomalous time windows. They are
prompted to return ranked leads using the same JSON schema as \sys, enabling
evaluation with a shared parser.
All baselines use the same underlying LLM (Gemini-3.1-Pro) as \sys
(\S\ref{sec:implementation}) and are evaluated on the same incidents. 

\PP{\sys configurations} \label{subsec:scc}
We evaluate  three variants of \sys.
All start from a bootstrapped library containing the starter operators,
additional operators synthesized by the LLM during bootstrap, and playbooks
over these operators. For the open-source benchmarks, we kept the starter
operators unchanged throughout the experiments.\footnote{Since
the DSB dataset was constructed by us,  there is a potential confounder where the  workload influences  design of the starter operators. The other benchmarks provide a complementary evaluation on
independent workloads.}
We refer to the three successive serving attempts as
Stages~1, 2, and~3.
\emph{\sys: stage 1 only} selects and instantiates an existing
playbook (Stage~1; Mode~1.A).
\emph{\sys: stages 1 + 2} additionally permits composition of
a new DSL program using existing operators when Stage~1 fails
(Stage~2; Mode~1.B).
Full \sys further permits synthesis of new operators and
composition of a program using the expanded library when
Stage~2 fails (Stage~3; level~2).

\PP{Evaluation metrics}
We use a two-dimensional metric, Accuracy@$k,i$, to evaluate iterative
RCA: the fraction of incidents where the true root cause appears among
the first $k$ ranked leads in at least one of up to $i$ attempts.
This captures both the number of leads an analyst must inspect per
attempt and the opportunities to revise a diagnosis.
We report Accuracy@3,3 on DSB and the three \openrca deployments,
allowing three ranked leads per attempt and three attempts total
(the initial attempt plus up to two retries). The Stage~1-only and
Stages~1+2 configurations correspond to Accuracy@3,1 and Accuracy@3,2,
respectively.
For ORCA-bench, we report Recall@5,3: for each window, we take the
highest fraction of labeled root causes found among the first five leads
in any of up to three attempts, then average across windows.
The Stage~1-only and Stages~1+2 configurations use Recall@5,1 and
Recall@5,2, respectively. On a retry, the
system is prompted to produce a different ranked list of root causes from
its previous attempt. Baselines repeat their investigation workflow, while
\sys progresses through playbook selection, composition using existing
operators, and evolution with newly synthesized operators.
We  report the total analysis cost, including LLM
inference and system costs.

\subsection{End-to-end evaluation}\label{sec:end-to-end}

\PP{Accuracy}
Figure~\ref{fig:market} compares \sys vs.\   three baselines across five
datasets. \sys matches or exceeds every
baseline on each dataset's metric. Its largest gain over the strongest
baseline for a dataset is $34.7$ percentage points
on DSB, or $62\%$ in relative terms. The strongest baseline
scores are $0.60$ for RCA-Agent on \market and $0.70$ and $0.80$ for
\terminus on \bank and \telecom.

Together, this validates  that the region explored by \sys on the
\eespectrum (Figure~\ref{fig:spectrum}) delivers  high  accuracy for
RCA tasks despite the restrictions on the program structure.
Since \terminus is the strongest baseline on four of the five datasets, we
use it for the incident-wise comparison in \S\ref{factor analysis} later.

\PP{Cost}
Figure~\ref{fig:market-cost} compares total analysis cost, including LLM
inference and system cost. \sys has the lowest cost on every
dataset, reducing cost by $5.4\times$ on average and up to $12.1\times$ on
DSB relative to the least expensive baseline on each dataset. 
Although \sys incurs higher system running 
cost,  the  savings in LLM inference cost more than offset this overhead. 

\PP{Recipe stage contributions}
Across the 70 incidents in \market, \bank, and \telecom,
32 ($45.7\%$) required Stage~2, and 19 ($27.1\%$) proceeded
to Stage~3. These stages correctly localized an additional
13 and 6 incidents, respectively, beyond the 38 handled
by Stage~1. Using the stage values in Figure~\ref{fig:market},
Stage~2 improves accuracy/recall over Stage~1 by $41.6\%$ on average
across all five datasets, averaging per-dataset relative gains.
Stage~3 improves accuracy/recall over Stage~2 by $7\%$ under the same
averaging convention; on \bank, this gain is $17\%$.
\begin{figure}[t]
    \centering
    \includegraphics[width=0.8\linewidth]{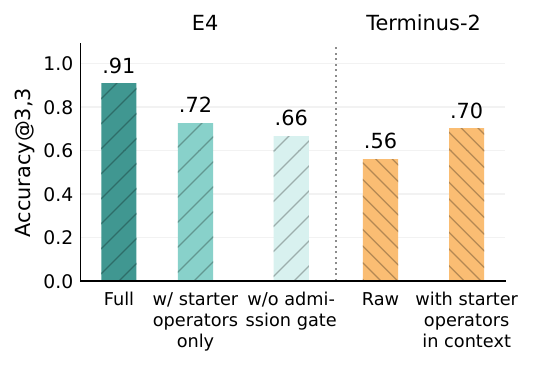}
    \vskip -0.2cm
    \compactcaption{Contribution of LLM-based operator synthesis and the admission
  gate in \sys for DSB dataset, compared with providing starter operators in \terminus's context.}
    \label{fig:factor-dsb}
\end{figure}

\PP{DSB incident types}
Figure~\ref{fig:acc} compares DSB incident types using the plotted
Accuracy@3,$i$ estimates. Averaging equally across fault types within each
group, \sys achieves $1.22\times$ the mean accuracy of \terminus on
client-side faults and $2.47\times$ on backend faults. These gains suggest
that \sys is robust across different fault types in both client-side
diagnosis and backend localization.

\subsection{Illustrative examples}
\label{sec:stage-contributions}

We illustrate findings from our evaluation datasets through examples
of how \sys prepares its operator library, reuses playbooks in
Stage~1, composes new analyses in Stage~2, and synthesizes new
operators in Stage~3.

\begin{packeditemize}

\item In \market, an incident flagged increased average latency
at the product-catalog service. In Stage~1, \sys selected an
existing metric-shift playbook, which compared component metrics
against historical baselines and combined evidence across abnormal
series. It ranked a shipping-service instance first and identified
excessive read I/O, matching the ground truth fault without
composing a new DAG or synthesizing an operator.

\item Another \market incident flagged increased average latency
at the shipping service. The metric-shift playbook selected in
Stage~1 did not identify the cause. In Stage~2, \sys composed a DAG
combining span-latency analysis with trace-flow attribution and
component-metric evidence. Using only existing operators, it
ranked a frontend instance first and identified network latency, matching the ground truth fault.

\item In \bank, an incident flagged increased average service
latency. Stages~1 and~2 did not yield a correct diagnosis among
the top three leads. \sys identified that its analyses measured
metric deviations without considering whether they tracked the
affected KPI. In Stage~3, it generated an operator that weights
metric anomaly scores by their temporal correlation with the KPI.
The resulting DAG ranked the affected application-server instance
first, matching the ground truth component.

\end{packeditemize}

\subsection{Ablation Study}\label{factor analysis}

\PP{LLM-based operator synthesis}
Figure~\ref{fig:factor-dsb}'s estimates decrease from $0.91$ for full \sys to $0.72$ with starter operators alone. This supports the design choice of extending the starter library using an LLM during bootstrapping (Section~\ref{sec:bootstrap}).

\PP{Admission gate}
Without the admission gate (\S\ref{sec:evolution}), accuracy falls to $0.66$. This suggests the value of retaining new operators only after they help resolve an incident.

\PP{Operators in context}
Providing starter operators in \terminus's context raises accuracy from $0.56$ to $0.70$, while full \sys reaches $0.91$. This suggests our choice of starter operators is generally useful, but leaves a gap to \sys's structured composition and evolution (Sections~\ref{sec:composition} and~\ref{sec:evolution}).

\begin{figure}[t]
    \centering
    \begin{subfigure}[t]{0.43\linewidth}
        \centering
        \includegraphics[width=\linewidth]{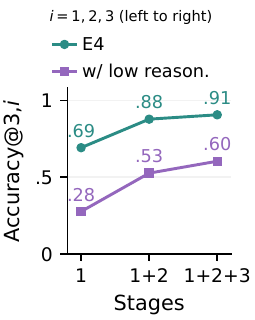}
        \caption{Reasoning capacity.}
        \label{fig:thinking}
    \end{subfigure}
    \hfill
    \begin{subfigure}[t]{0.43\linewidth}
        \centering
        \includegraphics[width=\linewidth]{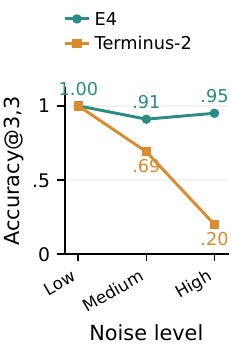}
        \caption{Impact of noise.}
        \label{fig:sensitivity-acc-size}
    \end{subfigure}
    \compactcaption{Accuracy for (a) low and high reasoning capacities over the three stages and (b) full \sys and \terminus at three dataset noise levels, both on the DSB dataset.}
    \label{fig:factor_analysis}
    \label{fig:sensitivity}
\end{figure}

\PP{Impact of LLM's reasoning power}
Figure~\ref{fig:thinking} compares the current DSB stage values with a low-reasoning curve. Accuracy@3,$i$ values of $0.28$, $0.53$, and $0.60$ for $i=1,2,3$, respectively, ending $33.3\%$ below full \sys hinting decent performance even at lower reasoning levels.

\PP{Robustness}
 We vary background noise by adding confounding activity, decoy
  spikes, and timing jitter. Figure~\ref{fig:sensitivity-acc-size}
  compares low, medium, and high noise. From low to high noise, \sys's accuracy drops by $5\%$, compared with $80\%$ for \terminus; under medium noise,
  their Accuracy@3,3 values are $0.91$ and $0.69$.
  Under high noise, Stage~2 improves accuracy over Stage~1 by
  $27\%$, suggesting that combining complementary evidence helps
  distinguish faults from background activity.

\subsection{Pilot study}
We evaluated \sys on 6 incidents sampled over two weeks from 
three large-scale production media-streaming services monitored by a large analytics
provider. The incidents involved login errors, session errors,
search success, and user abandonment. For each incident, we manually
established ground truth by identifying the affected user cohort from
concentrated complaints in application store reviews. \sys identified the correct
cohort in 5 of 6 incidents. Three without additional hints
and two with analyst hints guiding operator synthesis or DAG
refinement. Among these successful cases, the correct cohort ranked
first in three and second in two. We highlight two interesting incidents
\begin{itemize}
    \item 
In one incident, an app-store review reported that adding an item to the cart never completed. Investigating abandoned add-to-cart flows with errors, \sys successfully identified a third-party advertising endpoint as the primary root-cause lead. 
\item In another incident, an analyst asked which user actions preceded
  app crashes. \sys successfully discovered applying a coupon followed by selecting an installment-payment
  (EMI) option led to these crashes.
\end{itemize}

\section{Other related work}

We discussed some of the closest related work in \S\ref{rw1}.
Here we discuss other related work.

\PP{Observability and monitoring systems} 
RCA relies on diverse telemetry data and business-centric KPIs of interest. There are significant efforts in industry and academia on building such observability, monitoring, and, analytics systems (e.g.,~\cite{otel,datadog,jiang2016cfa,jaeger,lindon2022anytime,fu2021real}).  There are also many anomaly detection techniques built on top of these systems (e.g.,~\cite{xu2022anomaly, audibert2020usad, argos}).  These efforts are orthogonal to \sys and our system is flexible to accommodate telemetry data from these systems. 

\PP{Agentic systems for other aspects of operations} While we focused on troubleshooting, there are efforts underway to adopt agentic workflows in other aspects of system and network operations (e.g.,~\cite{wu2025leveraging,zhou2025meshagent,netarena}). These are orthogonal to our focus, but some of the work on configuration changes could be related both as causes of issues and remediations suggested after RCA. We evaluated works analogous to tool-augmented LLMs ~\cite{wang2024rcagent} in \S\ref{factor analysis}, however they were significantly outperformed by \sys.

\PP{Optimizing agents}
An interesting direction for future work is to employ agent optimization techniques~\cite{agrawal2025gepa,parthasarathi2025grpo,yao2022react,budgetaware,camdf}, including prompt optimization, to improve \sys.

\section{Discussion}
\label{sec:discussion}

\PP{Looking ahead as LLMs improve}
  We expect LLM agents to continue improving, and consider it plausible
  that
  generic agents will eventually outperform specialized harnesses in RCA
  accuracy.  For example, on DSB our direct LLM baseline improved from $8\%$ accuracy with
  Gemini-3-Pro to $28\%$ with Gemini-3.1-Pro (Figure~\ref{fig:market}); even at
  the latter accuracy it remains approximately $12\times$ more expensive than \sys. Production RCA will still require efficient execution,
  explainable results, and low operator effort. 
  \sys's
  analysis via a DAG of predefined operators avoids excessive token consumption during an incident and also enables reuse of existing
  pipelines that can be extended across successive investigations. This reduces repeated work while making the diagnosis explainable. As stronger LLMs improve the analysis accuracy, we believe \sys will still provide the mechanisms needed to
  turn
  that reasoning into an efficient and explainable operational system.

\PP{Data safety} Many enterprises may be wary of their data or workflows leaking to frontier models. \sys's design is advantageous in this scenario because the LLM never needs full telemetry dumps: it only sees the incident description, the operator catalog, and the dataset schema whereas the data is accessed locally by the execution engine via \sys's operators. 

\PP{Learning improvements}
 An interesting direction of future work is considering a continuously learning version of \sys or using ideas like test-time training~\cite{wang2025test}. That is, over time, \sys can build and refine  its library of  operators, DAG playbooks, and hypothesis  exploration prioritization for a given deployment. This could be further augmented with synthetic scenarios and/or human feedback as well.

\PP{\Structuredimagination for other Ops tasks} \sys's \structuredimagination is a promising approach to design agentic systems in various other troubleshooting domains e.g. ``grey failures'' such as silent packet drops (e.g.,~\cite{GrayFailure2017, Costa2022}), failures in the network layer (e.g.,~\cite{007,sherlock,netbouncer,maxCoverage,KDD-Fault-Localization,simon,netsonar,flock,netpoirot,deTector}), storage systems (e.g.,~\cite{deepview}), backend infrastructure (e.g.,~\cite{sage,murphy,netseer,explainIT,netmedic}) and large software systems (e.g.,~\cite{errLog,rept}). While we focused on RCA, the same idea could also be applied for anomaly detection. We leave these for future work.

\PP{Other uses of \structuredimagination} While we apply \structuredimagination to RCA/troubleshooting, the paradigm is more broadly valuable; prior work outlines its use for network configuration generation and red-teaming~\cite{sinha2026constrained}. The same paradigm with a different set of operators could also help with revenue debugging~\cite{adtributor}, aspects of exploratory data analysis and reliable information retrieval.

\section{Conclusion}
We presented \sys, an agentic RCA system built around \structuredimagination: the agent generates a response via a typed dataflow DAG based on domain specific operators for troubleshooting. This approach leads to the agent's results being accurate, explainable, verifiable and cost-effective. Across a mix of simulated and production workloads, \sys achieves up to $62\%$ better accuracy compared to state-of-the-art solutions, while providing more explainable responses at up to $12\times$ reduced cost. 

\medskip \noindent {\bf Ethics Statement:} Most of our experiments use real-world inspired incidents on a simulator. For the pilot study, we engaged domain experts inside an analytics provider and used real-world  data from one of this provider's customers. This does not contain any personally identifiable information and our analysis is in conformance with organizational policies.

\clearpage
\bibliographystyle{plain}
\bibliography{reference,rca}

\appendix
\onecolumn
{\centering\Large\bfseries APPENDIX\par}
\section{Dataset details} \label{app:dataset-details}

The \market, \bank, and \telecom datasets in \openrca originate from
the AIOps Challenge series~\cite{xu2025openrca}.

\PP{\market}
We use Market's cloudbed-1 deployment, an online-shopping microservice
system. Its telemetry combines distributed traces, service and proxy
logs, and metrics from services, containers, nodes, and the service mesh.
Service KPIs include request success rate, mean response time, and
request count. Faults include service failures, network problems,
and CPU, memory, and storage pressure.

\PP{\bank}
This dataset emulates a banking system with application servers,
databases, and caches. Its application KPIs and metrics are similar
to \market, including success rates, mean response times, and request
counts. Faults include resource saturation, network delay and loss,
and JVM failures.

\PP{\telecom}
This telecommunications system contains application services,
middleware, and databases. Telemetry includes distributed traces and
application, service, container, node, and middleware metrics. Service
KPIs include success rates, average response times, and request volumes.
Faults include CPU stress, network delay and loss, database connection
limits, and database shutdowns.

\PP{ORCA-bench}
ORCA-bench~\cite{gong2026orca} uses the Astronomy Shop microservice
application from the OpenTelemetry Demo~\cite{otel}. It contains six days
of metrics, logs, and traces, with frontend request error rates and
tail latency ($p_{99}$) as KPIs. Faults can be isolated or co-occurring,
including concurrent, cascading, and sequential failures. We construct
incident windows from frontend KPI anomalies without using
fault-injection times; a window may contain multiple root causes.

\PP{DSB}
Our extension of the DeathStarBench Social Network application combines
simulated client attributes, requests, and session-event histories with
backend traces and resource metrics. Client-side KPIs include workflow
completion rates and average durations; backend KPIs include request
error rates and average latency. Client-side faults affect cohorts,
URLs, or user behaviors, while backend faults introduce persistent or
intermittent network latency and CPU, memory, or disk stress.
Appendix~\ref{app:dsb-data} details the workload and fault-generation
procedures.

\PP{Incident counts}
Table~\ref{tab:datasets} counts the full benchmark corpora. For \openrca, we count distinct combinations of
timestamp, component, and fault reason in the original label files:
70 for Market's cloudbed-1, 136 for \bank, and 51 for \telecom.
For ORCA-bench, the original six-day schedule contains 36 continuous
fault episodes across 13 scenarios~\cite{gong2026orca}. Co-occurring
faults count separately; changes in severity during an active fault
remain part of the same episode. Repeated query variants and
no-fault controls do not add incidents. DSB contains 64 generated cases.

\PP{Telemetry sizes}
Uncompressed source telemetry totals 42.9, 11.9, 26.4, 17.7, and
25.7 decimal GB for DSB, \market, \bank, \telecom, and ORCA-bench,
respectively. Per-incident sizes in Table~\ref{tab:datasets} use compact JSON
inputs in decimal MB and report the mean and population standard
deviation. The size summaries cover 64 DSB inputs, 20 \market incidents,
30 \bank incidents, 20 \telecom incidents, and all 15 ORCA-bench windows,
including the two controls. DSB size measurements exclude metadata-only
files and sampled copies. Total source sizes and compact per-incident
sizes measure different representations and scopes.

\section{DSB dataset construction} \label{app:dsb-data}

\PP{Workload and collection}
The fault-injection harness drives the DeathStarBench Social Network backend
through simulated client sessions. Sessions carry client attributes and
histories and exercise timeline reads and post composition alongside generic
client-side URL flows. Each case records a historical window and an anomalous
window, followed by fault cleanup and trace collection. Workload settings vary
across the saved cases: 22 contain 1,200 sessions and 42 contain 4,800 sessions
across the two windows. The runner supports concurrent session generation,
with defaults of four workers and ten connections per worker.
It collects backend traces from Jaeger~\cite{jaeger} and exports per-container
CPU, memory, and disk metrics as spans through Jaeger's Zipkin
endpoint~\cite{zipkin}. The resulting artifact combines request records,
session events, traces, resource metrics, and window boundaries, with the
injected target recorded as ground truth.

\PP{Client-side injections}
Clients carry cohort attributes for country, state, city, ISP, operating
system, browser, and application version. For cohort faults, the harness
selects an attribute--value pair and applies the fault only to matching
clients. Cohort failures return HTTP 503 responses with a sampled probability
of 30--80\%; cohort-duration faults add 1--5~s to client-observed latency.
URL faults affect selected primary or backup URLs in the client flow.
History-dependent faults add latency to a client's second session or fail
requests following a selected sequence of events. These predicates require
the diagnosis to use session histories in addition to static client
attributes.

\PP{Backend injections}
Network faults use \texttt{tc netem} to introduce persistent or intermittent
latency at a target service. Resource faults use \texttt{stress-ng} to stress
CPU, memory, or disk. Background perturbations on other containers, including
confounding resource activity and decoy spikes, make elevated activity alone
insufficient to identify the injected target. The 64-case corpus contains
10 persistent-latency cases, four intermittent-latency cases, and six cases
for each resource fault. Its remaining 32 cases cover client-side faults:
four each for cohort duration and cohort failure, and six each for primary
URLs, backup URLs, second-session latency, and event sequences.

\section{DSL grammar} \label{app:dsl-grammar}

\PP{Preliminaries}
A deployment exposes \emph{telemetry} $T$: the set of tables its adapter
publishes, whose column names are $\mathrm{Schema}(T)$. Every incident
supplies two intervals over $T$, an \emph{anomalous window} and a
\emph{baseline window}. A \emph{candidate} names a possible root cause, such
as a client cohort given by an attribute--value pair, a backend service or
host, a URL, or an event pattern. A \emph{scored candidate} adds a score and
the evidence behind it, and a \emph{ranked answer} is the ordered list of
leads returned to the analyst.

The DSL is the pair $D = (G,O)$ of the fixed grammar $G$ below and an operator
vocabulary $O$. An \emph{operator} $o \in O$ is a typed function with a
declared signature, and its kind follows from its output type, so the grammar
needs no role annotations. $O$ is the vocabulary active for the deployment:
bootstrap specializes the starter library to $\mathrm{Schema}(T)$
(\S\ref{sec:bootstrap}), evolution adds synthesized operators, and eviction
removes unused ones (\S\ref{sec:evolution}). $O$ therefore changes between
incidents while $G$ does not. A \emph{program} is a straight-line sequence of
named bindings ending in one answer; a \emph{playbook} is a program whose
parameters are instantiated per incident. Programs are pure: they read only
$T$, the two windows, and their own earlier bindings.

\PP{Program structure}
The grammar writes the analysis DAG in topological order. Each production is
shown beside the conditions, in italics, that a program using it must satisfy.

\newcommand{\grammarrow}[2]{\par\noindent
  \begin{minipage}[t]{0.47\textwidth}\small
  \(\displaystyle\begin{array}[t]{@{}l@{}}#1\end{array}\)
  \end{minipage}\hfill
  \begin{minipage}[t]{0.5\textwidth}\small\itshape #2\end{minipage}\par\smallskip}

\grammarrow{
\mathit{program} ::= \mathit{binding};\ \dots;\ \mathit{binding};\ \mathit{answer}
}{every name is defined exactly once; every argument names an earlier binding; every
binding is consumed by a later binding or by the answer}
\grammarrow{
\mathit{binding} ::= \mathit{data} \mid \mathit{candidates} \mid \mathit{scored}
}{}
\grammarrow{
\mathit{data} ::= \mathit{data\ name} : \mathsf{Telemetry} \\
\qquad = \mathbf{load}[\mathit{params}](T) \\[2pt]
\quad \mid \mathit{data\ name} : \mathsf{Aggregate} \\
\qquad = \mathbf{aggregate}[\mathit{statistic};\ \mathit{params}] \\
\qquad (\mathit{data\ name},\ \mathit{window},\ \mathit{grouping})
}{one window per binding: comparing two windows is analysis, done by the
bindings below}
\grammarrow{
\mathit{candidates} ::= \mathit{candidate\ name} : \mathsf{CandidateSet} \\
\qquad = \mathit{operator}\langle\mathsf{CandidateSet}\rangle[\mathit{params}] \\
\qquad\quad (\mathit{arg},\ \dots)
}{each $\mathit{arg}$ is a data name or an earlier candidate name, so a
candidate set may be pruned or expanded before it is scored}
\grammarrow{
\mathit{scored} ::= \mathit{scored\ name}_m : \mathsf{ScoredCandidates} \\
\qquad = \mathit{operator}\langle\mathsf{ScoredCandidates}\rangle[\mathit{params}] \\
\qquad\quad (\mathit{arg},\ \dots)
}{each $\mathit{arg}$ is a data name, a candidate name or an earlier scored
name; candidate arguments are optional, but with $C_m$ the candidate names
this binding consumes and $N$ all those the program binds,
$\bigcup_m C_m = N$, so every candidate set is scored}
\grammarrow{
\mathit{answer} ::= \mathbf{return}\ \mathit{operator}\langle\mathsf{RankedAnswer}\rangle \\
\qquad [\mathit{params}](\mathit{scored\ name}_1,\ \dots,\ \mathit{scored\ name}_k)
}{the program's single terminator, not a binding; $k \geq 1$, and the answer
consumes every scored name no other binding consumes}

\PP{Parameters, operators, and types}
\par\smallskip\noindent
{\small
\begin{minipage}[t]{0.47\textwidth}
\(\displaystyle
\begin{array}[t]{@{}l@{}}
\mathit{window} ::= \mathbf{anomalous\ window} \mid \mathbf{baseline\ window} \\[3pt]
\mathit{attribute} ::= A \in \mathrm{Schema}(T) \\[3pt]
\mathit{grouping} ::= [\,] \mid [\mathit{attribute},\ \dots,\ \mathit{attribute}] \\[3pt]
\mathit{statistic} ::= \textsf{count} \mid \textsf{distinct} \mid \textsf{mean} \mid \textsf{sum} \\
\qquad \mid \textsf{quantiles} \mid \textsf{topk} \mid \textsf{min} \mid \textsf{max} \\[3pt]
\mathit{params} ::= [\,] \mid [\,p = v \mid p = \mathbf{auto},\ \dots\,]
\end{array}
\)
\end{minipage}\hfill
\begin{minipage}[t]{0.5\textwidth}
\(\displaystyle
\begin{array}[t]{@{}l@{}}
\mathit{operator}\langle\tau\rangle ::= o \in O \textit{ with declared output type } \tau \\[3pt]
\tau ::= \mathsf{Telemetry} \mid \mathsf{Aggregate} \mid \mathsf{CandidateSet} \\
\qquad \mid \mathsf{ScoredCandidates} \mid \mathsf{RankedAnswer} \\[3pt]
\mathit{datatype} ::= \tau \mid \mathsf{Window} \mid \mathsf{Cohort} \mid \mathsf{Candidate} \\
\qquad \mid \mathsf{Number} \mid \mathsf{Set}\langle\mathit{datatype}\rangle \\
\qquad \mid \mathit{datatype} \times \mathit{datatype}
\end{array}
\)
\end{minipage}
}\par\medskip

The \textbf{load} operator selects a telemetry table, and \textbf{aggregate}
computes a statistic over an attribute within one window, optionally grouped
by columns of $\mathrm{Schema}(T)$. A parameter $p$ takes either a literal
$v$ or \textbf{auto}, which the engine estimates from the baseline window at
execution time (\S\ref{sec:implementation}). The trailing dots in operator
calls stand for the further arguments an operator's signature declares: a
signature is a list of $\mathit{datatype}$s, and the verifier
(\S\ref{sec:composition}) checks every argument against it.
$\mathsf{CandidateSet}$, $\mathsf{ScoredCandidates}$, and
$\mathsf{RankedAnswer}$ denote candidate collections, candidates with scores,
and ranked leads, respectively.

\begin{figure}[p]
\centering
\includegraphics[width=\textwidth]{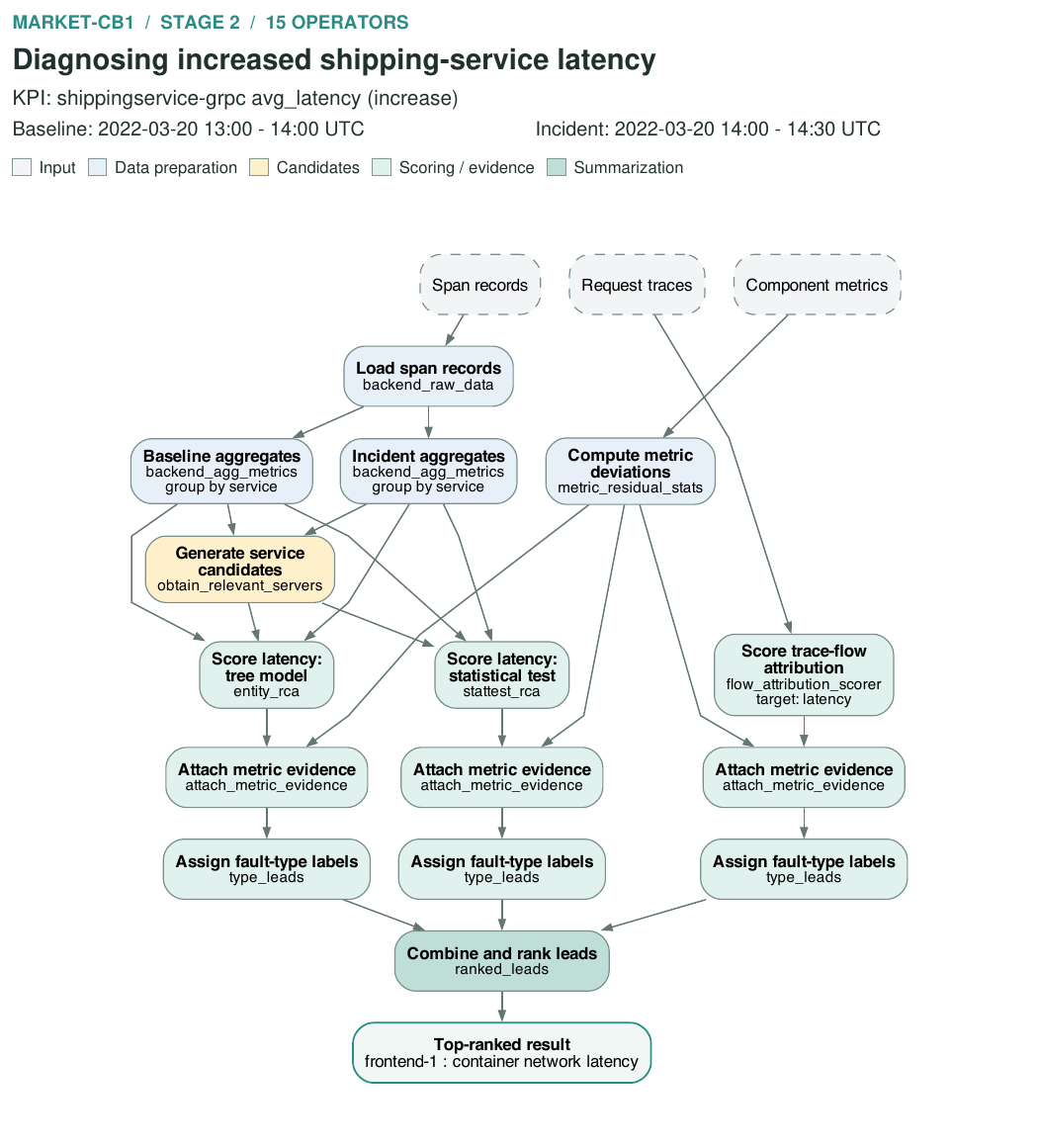}
\compactcaption{A saved Stage~2 analysis DAG for \market, with the observed
KPI, baseline and incident windows, and top-ranked result. All 15 operator
nodes and their data dependencies are shown. Shared aggregates and metric
deviations feed multiple analysis branches; literal parameters and
configuration inputs are omitted from the edges for readability.}
\label{fig:market-incident-dag}
\end{figure}

\PP{Example generated DAG}
Figure~\ref{fig:market-incident-dag} shows the complete 15-operator DAG
generated in Stage~2 for a \market incident. The alert concerned increased
shipping-service latency. The program shares baseline and incident
aggregates between two latency scorers, combines their results with
trace-flow attribution, and attaches component-metric evidence to all
three branches. Its top-ranked lead identifies a frontend instance with
network latency, matching the ground truth. All operators were already
available in the library; this attempt required composition only.

\section{Evolution as joint optimization} \label{app:objective}

Evolution (\S\ref{sec:evolution}) can be read as a joint optimization over
the program and the library, in the notation of
Appendix~\ref{app:dsl-grammar}. As in that section, $P$ denotes a
program and $L$ the library it is composed from. Since the grammar $G$ is
fixed, $L$ consists of the operator vocabulary $O$ and the playbooks written
over it. Let $P_B(i;L)$ be the program that stages (1) and (2) settle on for
incident $i$ within the attempt budget $B$ of Table~\ref{tab:params}, and
let $\mathrm{conf}(P)\in\{0,1\}$ be the analyst's verdict on the leads $P$
returns, $1$ if the analyst accepts them. For the distribution $Q$ of
incidents \sys observes, evolution seeks
\[
\underbrace{\min_{L\,\in\,\mathcal{A}}}_{\text{library}}\ \ \mathbb{E}_{i \sim Q}
\Big[\, 1 - \mathrm{conf}\big(\underbrace{P_B(i; L)}_{\text{program}}\big) \Big],
\]
where the expectation also covers the randomness of the LLM calls, and
$\mathcal{A}$ is the set of libraries the verifier admits: every playbook in
$L$ is well formed over $O$, and $|O|$ stays within the retention cap of
Table~\ref{tab:params}. Since $\mathrm{conf}$ is binary, the objective is the
probability that serving fails to satisfy the analyst.

Evolution solves this by modifying $P$ and $L$ alternately. Serving
(\S\ref{sec:runtime}) searches for a program $P$ over a fixed library $L$.
When it escalates on incident $i$, evolution changes $L$ by adding operators
and playbooks, and serving searches for $P$ again over the new library. The
change to $L$ is admitted only if serving then satisfies the analyst on $i$. Each change is
greedy: it is not revisited as later incidents arrive. Bootstrap
(\S\ref{sec:bootstrap}) only supplies the library this search starts from.

\section{Default settings} \label{app:settings}

Table~\ref{tab:params} lists the default runtime and data-processing
limits for \sys (\S\ref{sec:implementation}).

\begin{table}[!ht]
\centering
\small
\begin{tabular}[t]{@{}lr@{}}
\toprule
\textbf{Setting} & \textbf{Value} \\
\midrule
Nodes per program (maximum) & 40 \\
Stage~1 attempts (playbook selection) & 1 \\
Stage~2 attempts (composition) & 1 \\
Stage~3 attempts (after evolution) & 1 \\
Compile-time repairs per attempt & 2 \\
Feedback required to accept diagnosis & Satisfied \\
Leads retained per answer & 5 \\
Operators retained (with eviction) & 19 \\
New operators per evolution call (maximum) & 3 \\
\bottomrule
\end{tabular}\hspace{2.5em}%
\begin{tabular}[t]{@{}lr@{}}
\toprule
\textbf{Setting} & \textbf{Value} \\
\midrule
LLM temperature & 0.1 \\
Output tokens per call (maximum) & 65{,}535 \\
Thinking budget per call (tokens) & 8{,}000 \\
\midrule
Largest groups per digest aggregation & 8 \\
Time buckets per aggregation (maximum) & 500 \\
Characters per string (maximum) & 200 \\
Synthetic sessions per test case & 300 \\
Event-pattern length (maximum) & 3 \\
\bottomrule
\end{tabular}
\compactcaption{Default implementation settings. Accuracy evaluates the
first three ranked leads; ORCA-bench recall evaluates the first five.}
\label{tab:params}
\end{table}

\end{document}